\documentclass[draft]{agujournal2019}
\usepackage{apacite}
\usepackage{lineno}
\usepackage{amsmath}
\usepackage{algorithm}
\usepackage{dsfont}
\usepackage{algorithmicx}
\usepackage{algpseudocode}
\usepackage{tabularx}
\usepackage{booktabs}
\usepackage{url} 

\draftfalse

\journalname{Journal of Advances in Modeling Earth Systems (JAMES)}

\begin{document}

\title{Path-tracing Monte Carlo Library for 3D Radiative Transfer in Highly Resolved Cloudy Atmospheres}

\authors{
Najda Villefranque\affil{1,2},
Richard Fournier\affil{2},
Fleur Couvreux\affil{1},
St\'ephane Blanco\affil{2},
C\'eline Cornet\affil{3},
Vincent Eymet\affil{4},
Vincent Forest\affil{4},
Jean-Marc Tregan\affil{2}}

\affiliation{1}{Centre National de Recherches M\'et\'eorologiques (CNRM), UMR 3589 CNRS, M\'et\'eo France, Toulouse}
\affiliation{2}{Laboratoire Plasma et Conversion d'\'Energie (LAPLACE), UMR 5213 CNRS, Universit\'e Toulouse III}
\affiliation{3}{Univ. Lille, CNRS, UMR 8518 - LOA - Laboratoire d’Optique Atmosphérique, F-59000 Lille, France}
\affiliation{4}{M\'eso-Star, Toulouse, France}

\correspondingauthor{Najda Villefranque}{najda.villefranque@gmail.com}

\begin{keypoints}
\item Path-tracing library for flexible implementation of Monte Carlo algorithms in cloudy atmospheres
\item Null-collision algorithms and hierarchical grids to accelerate ray casting in large volumetric data
\item Insensitivity of radiative transfer computational cost to surface and volume complexity
\end{keypoints}

\justify

\begin{abstract}
  Interactions between clouds and radiation are at the root of many difficulties in numerically predicting future weather and climate and in retrieving the state of the atmosphere from remote sensing observations. The large range of issues related to these interactions, and in particular to three-dimensional interactions, motivated the development of accurate radiative tools able to compute all types of radiative metrics, from monochromatic, local and directional observables, to integrated energetic quantities. In the continuity of this community effort, we propose here an open-source library for general use in Monte Carlo algorithms. This library is devoted to the acceleration of path-tracing in complex data, typically high-resolution large-domain grounds and clouds. The main algorithmic advances embedded in the library are those related to the construction and traversal of hierarchical grids accelerating the tracing of paths through heterogeneous fields in null-collision (maximum cross-section) algorithms. We show that with these hierarchical grids, the computing time is only weakly sensitivive to the refinement of the volumetric data. The library is tested with a rendering algorithm that produces synthetic images of cloud radiances. Two other examples are given as illustrations, that are respectively used to analyse the transmission of solar radiation under a cloud together with its sensitivity to an optical parameter, and to assess a parametrization of 3D radiative effects of clouds.
\end{abstract}

\section{Introduction}
\label{part:introduction}
Radiative transfer, in the scope of atmospheric science, describes the propagation of radiation through a participating medium: the atmosphere, bounded by the Earth surface. Although many components of the Earth system interact with radiation, clouds play a key role because of their strong impact (globally cooling the Earth) \cite{ramanathan_cloud-radiative_1989}, their high frequency of occurrence \cite{rossow_international_2004} and their inherent complexity in both space and time \cite{davis_multifractal_1994}. Radiation and its interactions with clouds are involved in various atmospheric applications at a large range of scales: from the Earth energy balance and cycle relevant to numerical weather predictions 
\cite{hogan_radiation_2017} and climate studies \cite{cess_interpretation_1989,dufresne_assessment_2008}, to the inhomogeneous heating and cooling rates modifying dynamics and cloud processes at small scales \cite{klinger_effects_2017,klinger_cloud_2018,jakub_role_2017}, and to the retrieval of atmospheric state and properties from radiative quantities such as photon path statistics, spectrally resolved radiances or polarized reflectances \cite{cornet_cloud_2018}, observed by both active and passive remote sensors.

The three-dimensional (3D) models that have been previously developed in atmospheric science represent very accurately the interactions between clouds and radiation, but one-dimensional (1D) radiative transfer models are preferred in operational contexts, for their simplicity and efficiency. This is a demonstratedly poor approximation in cloudy conditions \cite{barker_assessing_2003,barker_estimation_2015}, particularly in broken cloud fields where cloud sides play an important role in the radiative fluxes distribution and divergence as they materialize a large portion of the interface between clouds and clear-air \cite{davies_effect_1978,harshvardhan_transport_1981,benner_three-dimensional_2001,pincus_accuracy_2005,hinkelman_effect_2007,kato_solar_2009}. A large-scale parametrization for 3D effects has recently been developed \cite{schafer_representing_2016,hogan_representing_2016}, leading to the very first estimation of the broadband, global, 3D radiative effect of clouds (around 2 W/m$^2$ after \citeA{schafer_what_2016}). This could not have occurred without the long-term efforts of a pioneering group of cloud-radiation scientists, who has been developing and using reference 3D radiative transfer models for the past fourty years, to analyze and document cloud-radiation 3D interactions (see \citeA{marshak_3d_2005,davis_solar_2010} and references therein). These 3D models can be divided into two categories: those using deterministic approaches (e.g. the Spherical Harmonics Discrete Ordinate Method \citeA{evans_spherical_1998}) and those using statistical approaches: Monte Carlo (MC) methods \cite{marchuk_elements_1980}. Our proposition builds upon one of the major strengths of MC models: the computing time being only weakly sensitive to the size of the geometrical and spectral dataset.

The theoretical reasons of this weak sensitivity have been identified since the origin of the method~(e.g. in \citeA{marshak_radiative_1995}), but it is only quite recently that Monte Carlo codes could practically handle highly refined surface descriptions and large cloud fields such as those produced by today's high-resolution atmopsheric models (with typically hundreds of million to a few billion grid points). This practicability has paved the way to numerous applications, even outside atmospheric sciences. The cinema industry has for instance recently started to make use of Monte Carlo for the physically-based rendering of cloudy scences~\cite{kutz_spectral_2017}. \citeA{brisc_physically_2019} have used a path-tracking physically-based software from the computer graphics community to render a video of a large-domain simulation produced by the ICON Large-Eddy Model (LEM) at 625m resolution. Also, since computing cost increases only linearly when adding integration dimensions (even for nonlinear processes, see~\citeA{dauchet_addressing_2018}), energetic engineers now consider combining solvers of cloud radiation and solvers of large scale energetic systems such as cities and solar plants into one single Monte Carlo algorithm~\cite{delatorre_monte_2014}. Altogether, observational, meteorological or climatic needs in atmospheric sciences, as well as similar requirements in other sciences, have motivated a community effort toward the practical handling of cloudy scenes of increasing size and increasing resolution. Along the line of the continuous development of Monte Carlo codes since the 1960s~\cite{collins_monte_1965, marchuk_monte_1980, marshak_radiative_1995, iwabuchi_modeling_2006, mayer_radiative_2009, pincus_computational_2009, cornet_three-dimensional_2010}, we here try to contribute with:
\begin{enumerate}
\item connections with the literature and practice of the computer graphics community,
\item a freely-available C library of general use in Monte Carlo problems involving large cloud scenes above complex surfaces. 
\end{enumerate}

Although we also present in this paper a code built with the library, we do not wish to focus on this particular code example, but on the library itself, that is designed to help the coding of a wide diversity of Monte Carlo algorithms while taking advantage of the recent developments in computer graphics. In today’s Monte Carlo codes, complexifying the ground description has no significant impact on the computing time. We show in this paper that, using the null-collision method (known as Maximum Cross Section in atmospheric science, \cite{marchuk_monte_1980}), together with computer science advances in the handling of large geometric data, computing time insensitivity can also be reached when increasing the cloud fields resolution.

Section~\ref{part:grilles} briefly recalls the principle of the acceleration grids that have been used to achieve the insensitivity of computing times to ground resolution and explains the reason why, until very recently, the same techniques could not be directly applied to volumes. Indeed, most Monte Carlo codes remain sensitive to the size or the refinement of the volume description because of the nonlinearity of Beer's extinction law. The end of this section is devoted to the well-established family of null-collision algorithms, here presented as a way to bypass this nonlinearity~\cite{galtier_integral_2013}, thus opening the door to acceleration grids for volumes also.

To the best of our knowledge the most advanced proposition along this line, in the field of cloud radiation, is in~\citeA{iwabuchi_multispectral_2017}. They use null-collision algorithms in acceleration grids, but in our sense, they do not implement all the possible benefits of acceleration structures: they do not show that they can lead to computing times that are insensitive to the resolution of the volumetric data. With distinct applicative objectives, strong efforts have also been made by the film industry, essentially by Disney Research, revisiting null-collision algorithms and turning them into a validated industrial practice~\cite{kutz_spectral_2017,novak_monte_2018,novak_residual_2014}.

Section~\ref{part:libraries} describes a new library inspired of such recent experiences. It allows the construction of acceleration grids for both surfaces and volumes. It makes use of the Embree library for surfaces~\cite{wald_embree_2014} and preserves its essence: an ensemble of low-level functions that help the design of Monte Carlo codes involving large geometric models and large volumetric datasets. The library elements remain independent, as much as possible, of the specificity of the (null-collision) Monte Carlo algorithm. 

In this sense, the present contribution is conceived in the spirit of the \emph{I3RC Community Monte Carlo model}~\cite{cahalan_i3rc:_2005, pincus_computational_2009, jones_design_2018}, designed as a platform to facilitate the development of atmospheric radiative transfer codes by radiation physicists in a wide range of applicative contexts. Another example of recent developments made in the form of a library grouping independent modules is RTE+RRTMGP (\textit{Radiative Transfer for Energetics + Rapid Radiative Transfer Model for GCMs, Parallel}, \citeA{pincus_balancing_2018}). Sharing their concerns on flexibility, replaceability and traceability, we have attached a strong attention to the abstractions we have used when splitting the library into elementary functions.

The algorithmic advances embedded in the library, that are at the heart of our proposition, are i/ the construction of hierarchical grids for both surfaces and volumes, and ii/ the filtering functions used as an abstraction to allow strict separation of the ray-casting procedure (iterating over the crossed voxels) from the Monte Carlo algorithm itself. It is demonstrated in Section~\ref{part:images} that the objective of computing time being insensitive to cloud field resolution is practically reached. This is illustrated using a rendering algorithm that produces synthetic images (fields of radiances) of scenes representing cloudy atmospheres, that we apply on a variety of cloud fields: stratocumulus, cumulus and congestus clouds. If this algorithm was designed to test the library in a concrete, challenging applicative context, the value of physically-based visualization of 3D atmospheric data in the assessment of model realism, process studies and inversion of satellite data is a clear motivation to our developments. 

As a perspective, two other radiative transfer algorithms are illustrated in Section~\ref{part:perspectives_other}. They were developed to study cloud-radiation interactions in the broader context of parametrization development, evaluation and calibration: the first algorithm evaluates ground fluxes together with their sensitivity to an optical parameter, and the second algorithm estimates the partition of ground fluxes into their direct and diffuse components. The state of the library and the current limitations at this stage are discussed in Section~\ref{part:perspectives_limitations}.

\section{Acceleration Grids for Large Surface and Volume Datasets}
\label{part:grilles}

\subsection{Why Monte Carlo Codes Can Be Insentitive to the Complexity of Ground Surfaces}

\begin{figure}\centering
	\includegraphics[scale=.6]{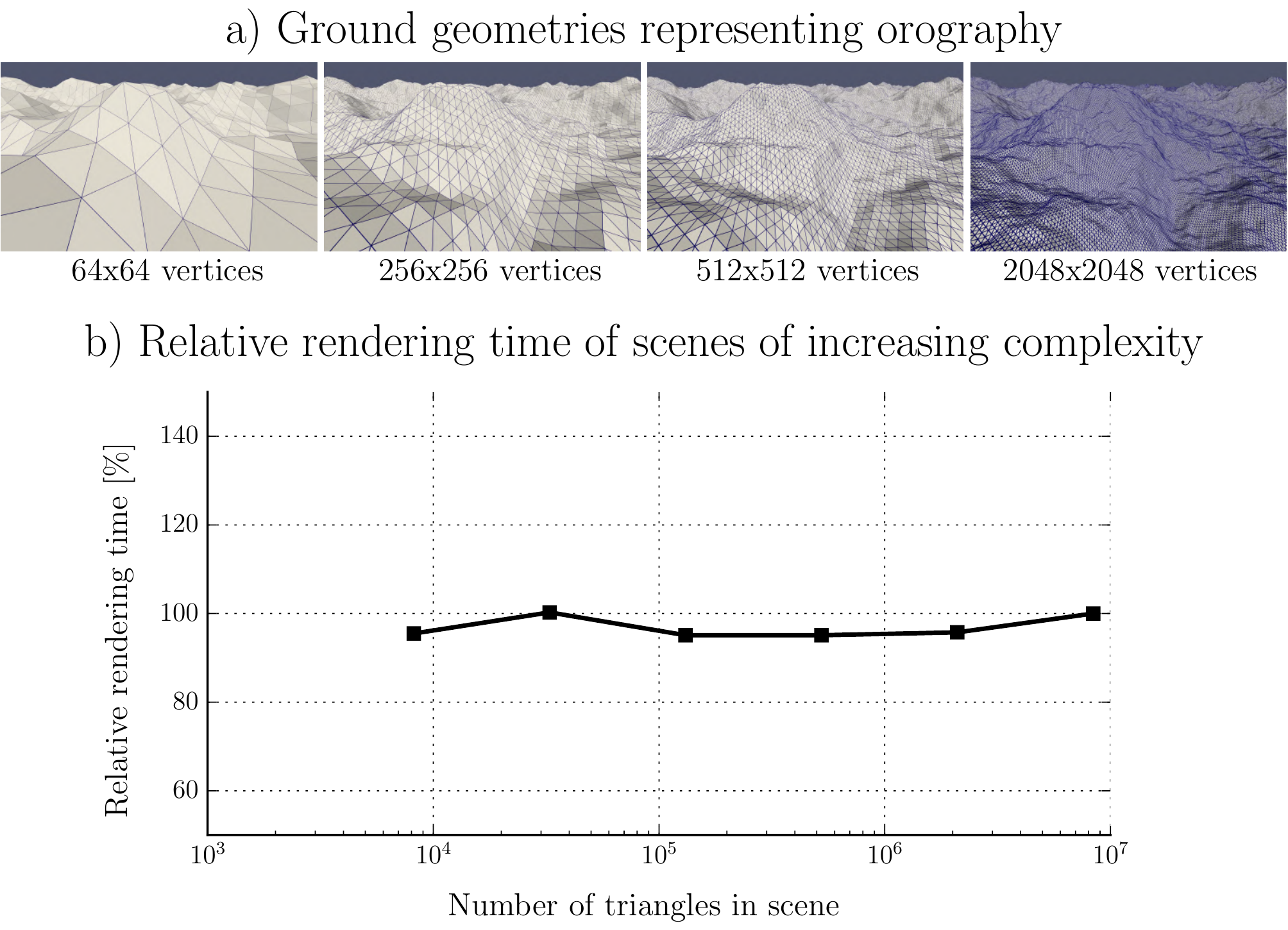}
  \caption{a) Surfaces representing orography described with an increasing number of triangles. b) Rendering time as a function of the number of triangles used to describe the surface, relatively to the rendering time of the scene using the most refined surface (2x2048x2048 triangles). The image using the most refined surface is shown in Figure~\ref{fig:imgs}-a). The same cloud scene is used for the other points on this plot, only the ground surface changes.}
  \label{fig:insensibilite_sol}
\end{figure}

Monte Carlo codes simulating radiation above a highly refined ground surface (discretized as millions of triangles) have to find the triangle that intersects the current ray, if any. This is a quite simple geometric problem, but speed requirements have motivated the development and use of acceleration structures to increase the efficency of ray-casting (see Appendix \ref{part:MC_surf} for a brief historical description). They change nothing to the surface geometry but organize the triangles in such a way that only those in the vicinity of the ray are checked for intersection. In practice, there is a precompuation phase in which the triangles are gathered into bounding boxes. When a ray is cast into the scene, the crossed bounding boxes are found and only the triangles inside them are tested for intersection. When dealing with large numbers of triangles, any such strategy reduces the computing time drastically by comparison with a systematic testing of all the triangles in the scene. But quite sophisticated acceleration structures were required before the cost of ray-casting procedures became fully independent of the number of trianges in the scene. Figure~\ref{fig:insensibilite_sol} illustrates this insensitivity of computing time to the complexity of the ground description. These sophisticated acceleration structures are hierarchical: they start with coarse bounding boxes that are recursively subdivided when they include too many triangles, allowing an adapted multi-level subdivision of space. Among the various available hierarchical grids, the choice is then made as a function of how much data need to be handle, whether these data fit in the available memory, the adopted parallelization or vectorization strategy, etc. This question is now very well documented and numerous libraries are available for fast implementation.  

\subsection{The Nonlinearity of Beer's Extinction Forbids the Straightforward Use of Acceleration Grids for Volumes}

\begin{figure}
\centering
\includegraphics[scale=1]{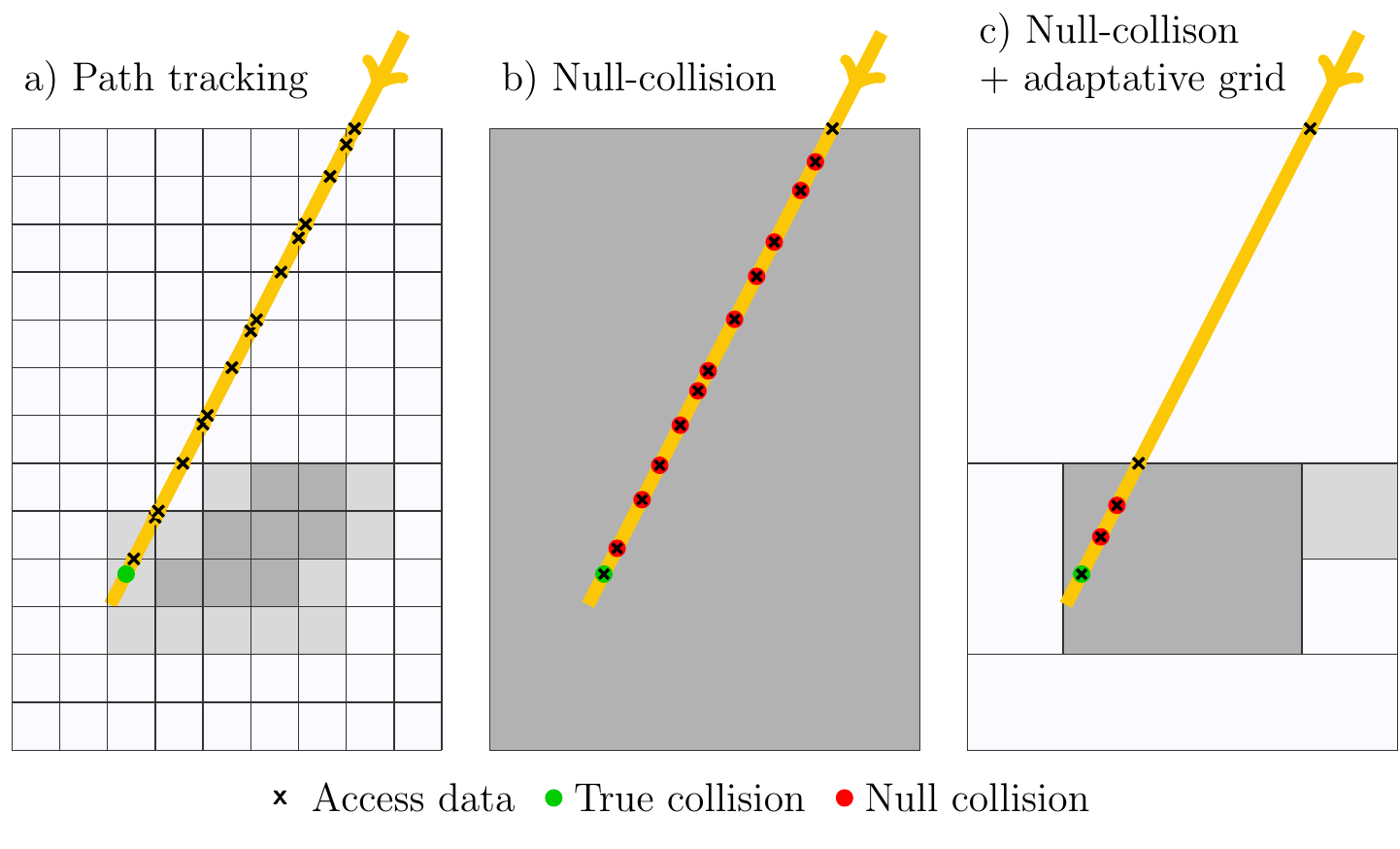}
        \caption{Two unbiased free-path sampling algorithms illustrated on a schematic 2D cloud field. Levels of gray represent the density of colliders in each cell. The thick yellow line represents a ray cast in the field. In both methods, data is accessed in each intersected cell. In \textit{path tracking} (a), the cost of the traversal is fully-dependent on the original data resolution. In \textit{null-collision} (b and c), coarser effective resolution is achieved by adding fictitious colliders in parts of the domain so as to make it homogeneous (b) or homogeneous-by-parts (c). The free-paths are sampled from the resulting modified field with two main consequences: i/ the effective density of colliders is overestimated in some parts of the domain, which is counterbalanced by rejecting some of the sampled collisions (yielding null collisions in red), and ii/ the cost of the traversal is decreased, and no longer depends on the original resolution. c) is a possible compromise between the two extreme strategies presented in a) and b).}
\label{fig:schema_ACN}
\end{figure}

When addressing the same question of handling large amounts of data, but now describing the state of the atmosphere, typically millions of elementary subvolumes in high-resolution discretisations used in Large Eddy Simulations (LES), an entirely new difficulty arises. Each ray will indeed successively cross ("intersect") several elementary volumes before finding the next volume collision location (absorption or scattering). Therefore, if the LES resolution is increased, the number of such successive crossings will increase proportionally (see Figure~\ref{fig:schema_ACN}-a).

Yet, the optical depth $\tau$ that is reconstructed by successively crossing the elementary volumes is nothing more than a one-dimensional integral of the extinction coefficient $k$ along the line of sight $s$. In the Monte Carlo context, evaluating such an integral should only imply that a distance $l_i$ be randomly sampled along the line of sight (e.g. uniformly):
\begin{equation}
        \tau_s =  \int_0^s k({\bf x}_l)\,\mathrm{d}l = \int_0^s \frac{\mathrm{d}l}{s} \left\{ k({\bf x}_l) \right\}s \approx \frac{1}{N} \sum_{i=1}^N k({\bf x}_{l_i})s
\end{equation}
The corresponding data-access difficulties would then reduce to finding in which elementary volumes lie the sampled ${\bf x}_{l_i}$ locations, and this could be achieved efficiently, like for surface intersections, by using acceleration grids that would organize the information about the spatial distribution of elementary volumes.

But this simple integral over the extinction coefficient (the optical thickness) cannot be statistically combined with the other integrals over photon-paths $\gamma$ (over scattering angles, over wavelengths, etc) in a straightforward manner because it appears inside the exponential of Beer's law that expresses the transmissivity ${\cal T}_\gamma$ along a path of length $s_\gamma$. The nonlinearity of the exponential imposes that either the optical thickness is evaluated in a deteministic way (abandoning the Monte Carlo approach for this part of the algorithm, i.e. crossing the elementary volumes successively and adding their contributions to evaluate $\tau_{s_\gamma}$ as in Figure~\ref{fig:schema_ACN}-a)) or a nonlinear Monte Carlo approach is used to handle simultaneously these two nonlinearly combined integrals. But until recently, the reported attempts to extend Monte Carlo to nonlinearly combined processes were scarce~\cite{dauchet_addressing_2018}. The deterministic approach was therefore commonly retained and path-tracing acceleration was difficult.

\subsection{Null-collision Algorithms and Their Integral Formulation Counterparts}
\label{part:acn-formulation}
A solution consists in adding virtual collisionners where true collisionners are scarce so that the total extinction coefficient is uniform throughout the whole field~\cite{woodcock_techniques_1965,marchuk_elements_1980}. This transforms the standard algorithms into new Monte Carlo algorithms, still unbiased (no approximation is introduced, \citeA{coleman_mathematical_1968}), where there is no integral inside the exponential any more. Of course, this is at the price of increasing the recursivity level of the path statistics (the number of successive scattering events in the modified paths $\hat{\gamma}$). Virtual collisionners have been added, hence when a photon encounters one of them, the collision must be rejected. This rejection takes the form of a purely forward scattering event, which is strictly equivalent to no scattering at all. This is illustrated in Figure~\ref{fig:schema_ACN}-b). These additional collision events can have a significant computational cost. There is therefore a compromise to be analysed: the multiple rejections of null-collision events may have a heavy computational cost, but nonlinearity vanishes from the integral form and efficient acceleration grids can be forseen, as illustrated in Figure~\ref{fig:schema_ACN}-c).

Before discussing these null-collision algorithms in terms of acceleration potentials, let us describe a first simple example: a null-collision Monte Carlo algorithm evaluating the direct monochromatic transmitted solar radiation at a location ${\bf x_0}$, through a cloudy atmosphere above a complex surface. The sun direction, ${\bf \omega}$, is computed from solar zenith and azimuth angles. We retain a backward algorithm in which the direct transmissivity $T({\bf x_0},{\bf \omega})$ is estimated by sampling $N$ radiative paths toward the sun, evaluating a path transmissivity $t$ for each path and taking the average: $T({\bf x_0},{\bf \omega})~\approx~\frac{1}{N}~\sum_{i=1}^N~t_i$. As per the null-collision approach, virtual collisionners defining a field of null-collision extinction coefficient $k_n$ are added such that the transformed medium, of extinction coefficient $\hat{k}~=~k~+~k_n$, is entirely homogeneous. Beer's law is used to sample the collision locations in the homogeneous $\hat{k}$-field. If no collision occurs before reaching the top of atmosphere, then $t = 1$. If a collision occurs at location ${\bf x_s}$, then the collision type is sampled. If the collision is a true collision, then $t = 0$. Otherwise the path is continued from ${\bf x_s}$ in a recursive manner. The resulting algorithm is the following:
\begin{enumerate}
        \item Set ${\bf x} = {\bf x_0}$.
        \item Cast a ray in the scene as if the volume was empty, originating from ${\bf x}$ in the direction ${\bf \omega}$, until either a surface is intersected or the ray reaches the top of the atmosphere (TOA).
	\item If a surface is intersected, return $t=0$ (the ground is opaque).
        \item If no surface is intersected, cast a ray in the homogeneous $\hat{k}$ volume: 
          \begin{enumerate}
          \item Compute $\hat{\tau}_{L} = \hat{k} L$ where $L$ is the distance from ${\bf x}$ up to the TOA in direction ${\bf \omega}$.
          \item Sample an optical thickness $\hat{\tau}_{s}$ according to Beer's extinction.
          \item If $\hat{\tau}_{s} > \hat{\tau}_{L}$, no collision is detected: return $t = 1$.
          \item If $\hat{\tau}_{s} < \hat{\tau}_{L}$, a collision is detected: set $s = \frac{ \hat{\tau}_{s} }{ \hat{k} }$, move to the collision location ${\bf x_s} = {\bf x} + s {\bf \omega}$ and access the local value $k({\bf x_s})$ of the field of extinction coefficient.
          \item Sample a random number $\epsilon$ uniformly in the unit interval in order to decide between a true and a null collision.
          \item If $\epsilon < \frac{k({\bf x_s})}{\hat{k}}$ the collision is true: return $t = 0$.
          \item If $\epsilon > \frac{k({\bf x_s})}{\hat{k}}$ the collision is null: go to 5.
          \end{enumerate}
        \item Set ${\bf x} = {\bf x_s}$ and loop to step 4.
\end{enumerate}
This algorithm has the following rigorous counterpart in terms of integral formulation (writing $T({\bf x_0},{\bf \omega})$ as an expectation~\cite{eymet_boundary-based_2005,dauchet_practice_2013,delatorre_monte_2014}):
\begin{eqnarray}
  \label{eq:intform}
        T({\bf x_0},{\bf \omega})&=& \int_0^\infty  \overbrace{\mathrm{d}\hat{\tau}_{s} \,  \exp{\big(-\hat{\tau}_{s}\big)}}^{(b)} \bigg(\overbrace{\mathcal{H}(\hat{\tau}_{s}-\hat{\tau}_{L})\big\{ 1\big\}}^{(c)} \quad  \nonumber \\ && + \underbrace{\mathcal{H}(\hat{\tau}_{L}-\hat{\tau}_{s})}_{(d)}\Big(\underbrace{\frac{k}{\hat{k}}\big\{0\big\}}_{(f)} + \underbrace{\big(1- \frac{k}{\hat{k}}\big)\big\{ T({\bf x_s},{\bf \omega}) \big\}}_{(g)} \Big) \bigg)
\end{eqnarray}
where $\mathcal{H}$ is the Heaviside function. Braces indicate correspondance with the steps described above in order to highlight the one-to-one equivalence between the formulation and the algorithm. This is highlighted here to explain one of our leading objectives when designing the library: facilitating a back and forth practice from one of these view points to the other, i.e. designing an algorithm by working on the integral formulation and analysing/modifying an existing algorithm by translating it into an integral expression (the expectation of the Monte Carlo estimator).

A typical example of such a practice is the question of evaluating the sensitivity of radiative metrics to uncertain optical parameters, with implications for data assimilation, atmospheric state retrievals, and analysis of the (3D) interactions between radiation and atmospheric or surface properties. The starting point is an existing Monte Carlo algorithm, that evaluates a given metric, e.g. the direct transmissivity $T({\bf x_0},{\bf \omega})$ in the above example. The objective is to transform the algorithm so that it also evaluates the derivative 
$\partial_\pi T({\bf x_0},{\bf \omega})$ with respect to a parameter $\pi$. The corresponding steps are
\begin{enumerate}
\item translating the algorithm into its integral counterpart (Equation~\eqref{eq:intform}),
\item derivating this integral with respect to $\pi$ and transforming it so as to retrieve the probability density functions (the paths) that were sampled in the original algorithm,
\begin{eqnarray}
  \label{eq:intform-sensib}
	\partial_\pi T({\bf x_0},{\bf \omega}) &=& \int_0^\infty  \mathrm{d}\hat{\tau}_{s} \,  \exp{\big(-\hat{\tau}_{s}\big)} \Bigg(\mathcal{H}(\hat{\tau}_{s}-\hat{\tau}_{L})\big\{0\big\} \quad \nonumber \\  && + \mathcal{H}(\hat{\tau}_{L}-\hat{\tau}_{s}) \Bigg(\frac{k}{\hat{k}}\big\{0\big\} + \big(1- \frac{k}{\hat{k}}\big) \nonumber \\ &&\times \Bigg\{  - \frac{\partial_\pi k({\bf x_s})}{\hat{k}({\bf x_s}) - k({\bf x_s})} T({\bf x_s},{\bf \omega}) + \partial_\pi T({\bf x_s},{\bf \omega}) \Bigg\} \Bigg) \Bigg)
\end{eqnarray}
\item translating the integral back to the algorithm, which here simply means that a new variable $\sigma$ is introduced that stores, at each null collision, the logarithmic derivative of the null-collision probability ($\sigma \leftarrow \sigma  - \frac{\partial_\pi k({\bf x_s})}{\hat{k}({\bf x_s}) - k({\bf x_s})}$), and that a Monte Carlo weight $t_\pi = \sigma \ t$ is outputted together with $t$. The sensitivity estimate is then the average of $t_\pi$ for the $N$ sampled paths: $\partial_\pi T({\bf x_0},{\bf \omega}) \approx \frac{1}{N} \sum_{i=1}^N t_{\pi,i}$
\end{enumerate}

In this presentation of null-collision algorithms and their integral formulations, two main features can be highlighted. First, in Equations~\eqref{eq:intform}~and~\eqref{eq:intform-sensib} the integral of $k$ along the line of sight, $\int_0^L k({\bf x_s}) ds$, does not appear inside the exponential anymore. Second, when working on the integral formulation (e.g. for sensitivities, an example of such simulation will be presented in \ref{subpart:sensib2}) new quantities may have to be computed and stored at each null collision (here, $\sigma$), which expands the practical significance of null-collision algorithms beyond simple rejection algorithms. This required close attention when designing the library.

\subsection{The Expected Features of Acceleration Grids for Path-tracing in Null-collision Algorithms} 
\label{part:grilles_expectations}

Such null-collision algorithms have been known since the origin of Monte Carlo in all fields of particle transport physics (under the name \emph{Maximum Cross-section} in atmospheric radiation~\cite{marchuk_elements_1980}), but they have essentially been considered as a trick to avoid the heavy coding of crossing elementary volumes one after the other. It is only very recently that they were theoretically analysed as a way to bypass the difficulties associated to the nonlinearity of Beer's extinction and to integrate the heterogeneities of $k$ along the path as part of the Monte Carlo integration itself. Among the first illustrated consequences of this revised viewpoint is the fact that acceleration grids could indeed be introduced for volumes~\cite{iwabuchi_multispectral_2017, kutz_spectral_2017, novak_monte_2018}. Three distinct objectives orient the design of such acceleration grids:
\begin{enumerate}
\item they should help adjusting the $\hat{k}$-field locally to minimize the computational cost of rejecting too many null collisions,
\item they should accelerate the traversal of the $\hat{k}$-field and allow fast access to the true $k$ value (the true atmospheric data) when a collision is found in the transformed field,
\item the precomputation cost associated to their construction must be small. 
\end{enumerate}
Indeed, it is not required that null-collisionners be added until the whole field of the extinction coefficient is uniform. The only requirement is that the spatial variations of $\hat{k}$ be simple enough to allow a fast sampling of the next collision location. Of course, if $\hat{k}$ is uniform the sampling is easy because the distribution is a simple exponential (Beer's extinction in a uniform field), but the sampling is also very simple if $\hat{k}$ is uniform by parts. So the acceleration grid will introduce voxels (super-cells in \citeA{iwabuchi_multispectral_2017}) and $\hat{k}$ will be uniform inside each voxel. The voxels will be chosen so that ${k_n}$ be as small as possible and only few null-collisions are introduced, therefore minimizing the computation time devoted to their rejection (see Figure~\ref{fig:schema_ACN}-c)). 

The second objective requires that the structure of the multi-scale grid be such that its traversal is fast and that when a given level is reached, accessing the corresponding data (the true field) is efficient. This is the same question as when accelerating the intersection with surfaces and the same algorithmic solutions can be used, mainly the use of hierarchical structures. As for surfaces, the grid should be refined as a function of collisionner density. This means that we have to moderate the statement that $\hat{k}$ should be as close to $k$ as possible: if we want $\hat{k}$ to match $k$ very closely, then the acceleration grid will be very refined (ideally as refined as the original field) and traversing the acceleration grid will be as expensive as computing the optical thickness deterministically. A compromise needs to be found and obviously this compromise is related to optical thickness. There is indeed no reason for $\hat{k}$ to match $k$ closely if the corresponding optical thickness is small and therefore little collisions will occur. As shown later in Section~\ref{part:images_tau}, an optical thickness of 1 to 10 appears to be a good compromise between voxel intersections and null collision rejections.

\section{A Path-tracing Library}
\label{part:libraries}

Section \ref{part:grilles} has stated that null-collision algorithms can be seen as a way to bypass the nonlinearity of Beer's extinction law, thus making it possible to develop acceleration strategies to cast rays into volumes, while benefiting from similar developments made for surface treatment in computer graphics. This section describes the path-tracing library that is at the heart of our proposition, explains how hierarchical grids can be constructed using the library (\ref{part:liblib}), and how the specificity of the ray-casting procedure implemented in the library allows the flexibility that physicists require when coding algorithms derived from integral formulations (\ref{part:libfiltre}).

\begin{figure}\centering
  \includegraphics[scale=1]{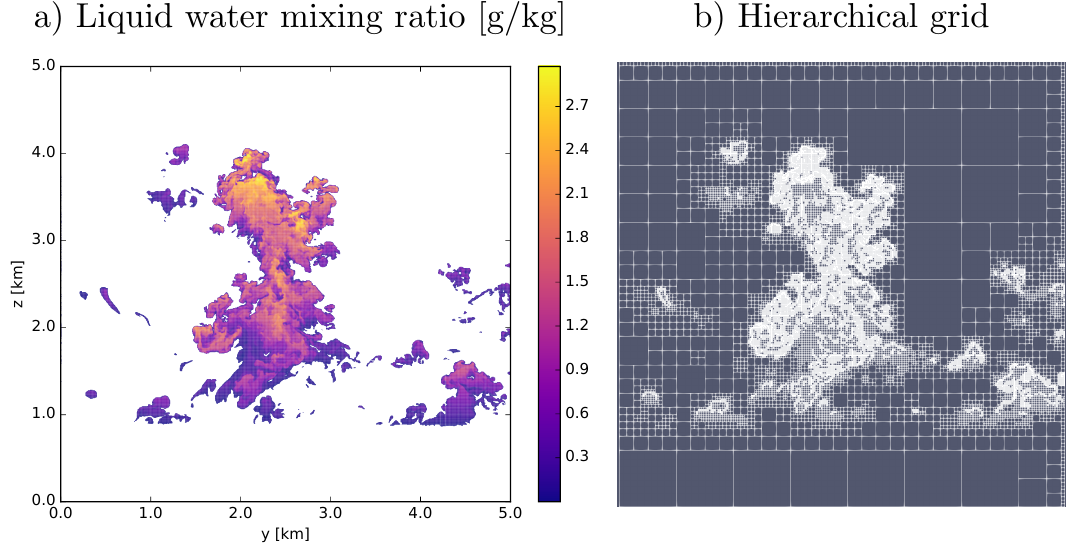}
        \caption{Vertical cross sections of a) liquid water mixing ratio from a highly resolved heterogeneous cloud field from a Large Eddy Simulation, and b) the hierarchical grid that was built from it. The original data is 38 Go in netCDF format, while the acceleration grid is 7.4 Go in VTK format.}
  \label{fig:hierarchical_structure}
\end{figure}

As mentionned before, the principal expected benefit of using null-collision algorithms in combination with acceleration grids is that the computing time dedicated to finding the location of next ray-medium interaction should no longer be dependent of the resolution of the input data. As an illustration of the data that are typically output from high-resolution atmospheric models run on large domains, Figure \ref{fig:hierarchical_structure}-a) shows a vertical cross section of liquid water mixing ratio in a highly refined cloud field that was produced by the Meso-NH \cite{lafore_meso-NH_1997,lac_overview_2018} Large Eddy Model, with 5 meters resolution in all three directions, on a 5$\times$5$\times$5 km$^3$ domain. The initial conditions and model set-up for this simulation (but with a 50 m resolution) is described in~\citeA{strauss_evaluation_2019}. The 3D fields of liquid and vapour water, temperature and pressure are partitioned into regular grids of 1000$^3$ cells, which represents about 38 Go of data. To these physical 3D fields, a spectral dimension issued from a k-distribution model \cite{mlawer_radiative_1997,iacono_radiative_2008} is added, that multiplies the amount of data by the thirty quadrature points used in the visible part of the solar spectrum. Details on the production of the physical data and the optical properties of cloud droplets and gas are presented in Appendix~\ref{app:data}. 

As many of the grid cells are ``clear" in most simulated 3D cloud fields, and thus hardly contribute to the total optical depth of the cloud scene, benefits of using null-collision algorithms and acceleration structures are expected to be important. In \citeA{iwabuchi_multispectral_2017}, a first step in the hierarchical treatment of these clear cells consists in separating the cloudy layer from the clear layers that stand above and below, and then generating acceleration grids at fixed resolutions that differ in clear and cloudy layers. Here, we show that we can go one step further by generating acceleration grids that, by their hierarchical nature, handle all the scales of horizontal and vertical variations of the extinction field. This is illustrated in Figure \ref{fig:hierarchical_structure}-b), that represents a cross section of the 3D acceleration grid constructed from the 3D 5m-resolution cloud field of Figure \ref{fig:hierarchical_structure}-a).

\subsection{Construction and Use of Hierarchical Grids}
\label{part:liblib}
A development environment constituted by a set of independent libraries is freely available online \cite{starengine}: the functions that are dedicated to the construction and crossing of hierarchical grids are implemented in Star-VoXel, which is one of these distributed libraries. These tools are addressed to radiative transfer specialists who are either developing new Monte Carlo codes or upgrading the ray-casting routines in existing ones.

Independent modules offering functionalities such as random sampling of probability density functions, parallel integration of a realization function, sampling and evaluation of scattering and reflection functions, and ray casting in surfaces and volumes are described in Table \ref{tab:modules} of Appendix \ref{app:lib}. The module that handles ray casting in surfaces is based on the Embree library \cite{wald_embree_2014}, that is the common standard in computer graphics. For volumes however, although solutions to render complex volumes exist for production purposes (a renderer based on the OpenVDB library \cite{museth_VDB_2013} was recently applied to a high-resolution 4D convective cloud field \cite{brisc_physically_2019}), it is our understanding that the management of volumetric data has not yet reached the same level of maturity as for surface data.

\subsubsection{Construction}
In our library, we chose to implement one specific type of {acceleration structure}: octrees, that are {hierarchical grids} that partition 3D data.  To construct these hierarchical grids, groups of $2^3$ cells containing the data (e.g. extinction coefficients) are recursively tested for merging. Since strategies for merging voxels control the balance between the cost of crossing the grid and the cost of rejecting null collisions, they should be thought in coherence with the specificity of the implemented algorithm. This is why no assumption on the input data, the merging strategy or the data that will be stored in the acceleration grid is made at the library level: it is left to the entire responsibility of the physicist. 

To build the hierarchical grid illustrated in Figure \ref{fig:hierarchical_structure}, an optical depth criterion is used to merge voxels that contain local extinction coefficients. The minimum and maximum extinction coefficients of the merged region are stored in each merged voxel. To handle the spectral dimension we build one octree per quadrature point, which is an arbitrary choice based on simplicity and that should be improved in future work. 

\subsubsection{Storage}
Since the paths will be tracked in the hierarchical grids, it is no longer required that the raw data fit into the main memory. The original input data are stored on disk and loaded into memory whenever a collision is found and its nature needs to be tested. The immediate benefit is that calculations in large cloud fields that would not fit into memory are now possible. Of course, time is then spent in loading / unloading chunks of data (fragments of contiguous data in memory or disk space) into / from the main memory which rapidly becomes prohibitive in terms of computational effort. Another limitation in the handling of huge volumetric data is that building octrees with a coarser (suboptimal) refinement might prove necessary since, as of now, the octrees are still stored into the main memory. 

However, strategies to improve performances have been anticipated in the library implementation. The library registers the voxels in a Morton order that preserves the spatial coherence of the 3D data in memory or on disk \cite{baert_out-of-core_2013}. The data are fragmented into fixed-size memory blocks \cite{laine_efficient_2010}, that can be efficiently (un)loaded by the operating system to handle out-of-core data \cite{tu_etree_2003}. This insures that whenever a ray interacts with several voxels in a limited spatial region, the relevant data are available in memory as of the first interaction that necessitated the loading of the corresponding data chunk.

\subsubsection{Crossing}
The last important functionality implemented in the library is the crossing of the hierarchical grid. The ray-casting procedure can be seen as a sophisticated ``do while loop": it is an abstract procedure that iterates in an ordered fashion on the voxels that are intersected by the ray. At each intersection, a filtering function (the ``loop body") is called. No assumption on either the nature of the data contained in the voxels, or on the treatment that will be applied by the filtering function upon voxel intersection, is made at the library level: again, it is left to the responsibility of the physicist. Exposing the physically-based motivations behind this choice of abstraction, materialized by the effective independence between ray casting and intersection treatment, is the object of the next subsection.

\subsection{Integral Formulations and Filtering Functions}
\label{part:libfiltre}
As mentioned before, a strong attention was devoted to the separation of concepts while designing the library. As much as possible, we tried to preserve a coherence with the computer graphics libraries from which we started~\cite{pharr_physically_2018, wald_embree_2014}, but above all we systematically favored all possible connections with the integral formulation concepts of the radiative transfer community.

The specificities of these formulations were illustrated when null-collision methods were introduced in Section \ref{part:acn-formulation}. One of these specificities is the recursivity associated to the rejection of null collision events. In order to separate the physical part of the code (e.g. where the treatment of scattering or reflection events is implemented) from the treatment associated to the recursivity of the ray casting, filtering functions are used. The same concept was introduced by the computer graphics community in order to deal with surface impacts that require a specific treatment inside the ray-casting function itself, for instance filtering out (ignoring) the ray intersections with transparent surfaces. In volumes, the objective is that the ray-casting procedure should not be exited at each crossed voxel, but only when a (true) collision is found. To that end, a filtering function is called at each voxel intersection: it at least handles null-collision rejections, but more sophisticated treatment might be needed depending on the algorithm. 

This implies that the filtering function is designed to be accessed by the physicist while implementing any algorithm. An example of a more complex requirement than rejecting null-collisions, that is bound to be handled by the filtering function, was illustrated with the example of evaluating sensitivities in Section~\ref{part:acn-formulation}. Treating intersected voxels by filtering them and optionally evaluating quantities at each intersection can be associated to other types of algorithmic operations due to transformations made at the integral formulation level. Sensitivity evaluation is only one example of such. A second example is the possibility of handling negative null-collision coefficients, i.e. configurations for which $\hat{k}$ is not a true overestimate of $k$~\cite{galtier_integral_2013}. A third example is the sampling of absorption lines when the gaseous part of $k$ cannot be precomputed in line-by-line Monte Carlo algorithms dealing with large spectroscopic databases~\cite{galtier_radiative_2016}. More generally speaking, as soon as the introduction of null collisions is perceived as a formal way to handle the nonlinearity of Beer's extinction in heterogeneous fields, the door is open to the design of Monte Carlo algorithms departing widely from the intuitive addition of virtual collisionners, and the use of filtering functions is a practical way to simplify such developments: the iteration over intersected voxels is handled by the ray-casting procedure, and the treatment specific to the recursive algorithm in its integral form can be directly implemented in the filtering function.

\section{Implementation and Performance Tests}
\label{part:images}
In this section, a rendering algorithm is implemented using the library described above, to show that null-collision algorithms that track paths in hierarchical structures allow to compute radiance fields of clouds described by large datasets (up to 1000$\times$1000$\times$1000 cells), and that the rendering time is almost insensitive to the resolution of the cloud field (i.e. to the size of the dataset). This is the main achievement reported in this paper, and this whole section is dedicated to the analysis of performances in terms of rendering time, as a function of the amount of volumetric data, but also of the type of clouds, and of the merging strategy used when constructing the hierarchical structures.

\subsection{The Algorithm}
Rendering images of highly resolved clouds is challenging in term of computational resources, yet 3D visualization of atmospheric data helps judging the realism of high-resolution simulations and provides information on the 3D paths of light and their interactions with clouds. Such rendering algorithms are also useful to evaluate the inversion procedures used to retrieve cloud parameters from satellites images. To render a virtual cloud scene, a virtual camera is positioned anywhere in 3D space and its position, target point and field-of-view define an image plane, that is discretized into a given number of square pixels. For each pixel, three independent Monte Carlo simulations are run to estimate the radiance that is incident at the camera, integrated over the small field-of-view defined by the pixel size, and integrated over the solar spectrum weighted by the responsitivity spectra of the three types of human eye cone cells \cite{smith_CIE_1931}. Pixels are distributed among the different nodes and threads whenever parallelization is active.

\begin{figure}\centering 
	\includegraphics[scale=.7]{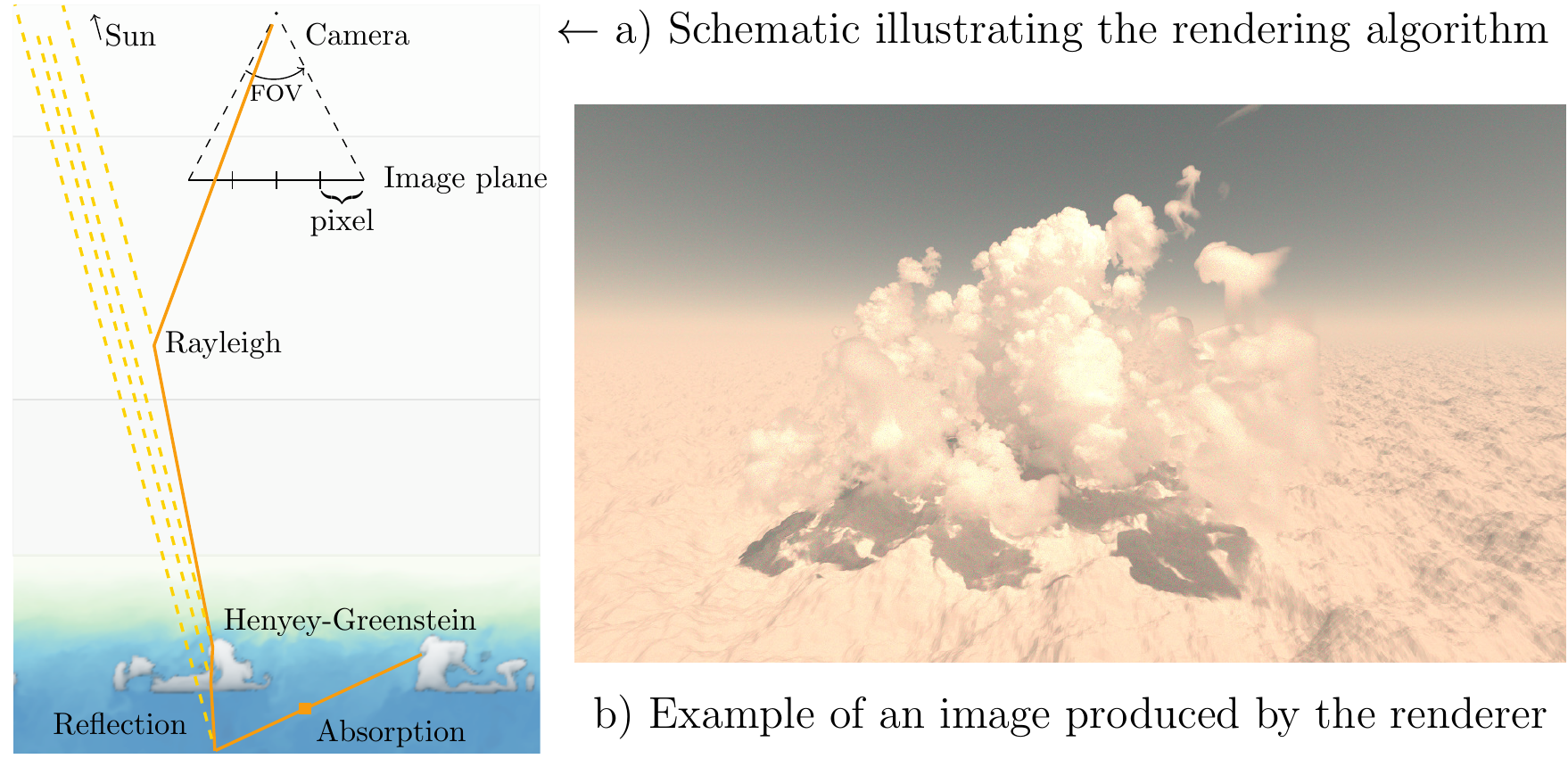}
        \caption{a) Schematic illustrating the rendering algorithm. The paths are tracked from a virtual camera throughout the medium until escape or absorption. At each interaction with the medium, the contribution of the direct sun, transmitted along the tracked path, is added to the path weight, as per the local estimate method, in a backward version. b) image of a high-resolution congestus cloud \cite{strauss_evaluation_2019} over a complex ground rendered with 4096 paths computed for each of the 3 spectral components of each of the 1280x720 pixels (11 324 620 800 paths in total). The camera and sun set-up is described in Table~\ref{table:scenes} in Appendix~\ref{app:lib}.}
	\label{fig:schema_path_image}
\end{figure}

The retained backward algorithm is as follows: paths are initiated at the camera. A direction $\omega$ is sampled in the solid angle defined by the pixel size and position in the image plane. A wavelength is sampled following the responsivity spectra of the current component. The narrow band in which lies the sampled wavelength is found in the k-distribution data. A quadrature point is sampled in the narrow band. The contribution of the direct sun is computed: if the current direction of propagation $\omega$ lies into the solar cone, and no surface intersection is found along the ray trajectory, then the ray is cast into the volume to compute the direct sun transmissivity, as per the algorithm described in \ref{part:acn-formulation}, but using in addition a variance reduction technique called \textit{decomposition tracking} \cite{novak_residual_2014,kutz_spectral_2017}. Otherwise, the direct contribution is null. Then, the path is tracked in the (null-collision) scattering medium to compute the contribution of the diffuse sun. Direct transmissivity between each two reflection or scattering events is evaluated in the absorbing volume and cumulated along the path. When the ray hits a surface the reflectivity of the ground is recovered and termination of the path is sampled accordingly. When a scattering event occurs, local scattering coefficients of the gas mixture and the cloud droplets are recovered, and the specie responsible for the scattering is sampled accordingly. Then, the surface or volume event is treated by sampling a new direction of propagation, following the appropriate scattering function (Henyey Greenstein for cloud droplets, Rayleigh for gas molecules, Lambertian for surfaces), and the ray is cast again in this new direction. The Henyey Greenstein phase function is used with asymetry parameter and single scattering albedo issued from Mie computations, at the wavelength lying at the center of the narrow band. The path is terminated when reaching the TOA or upon absorption by the ground or the volume (if the direct transmissivity between two events is null). Following the local estimate method \cite{marchuk_monte_1980, mayer_radiative_2009}, the path weight is updated at each surface and volume event by adding the sun direct transmissivity from the TOA to the event location, weighted by the probability of reflection or scattering from the sun direction into the tracked direction, and by the transmissivity along the tracked path from the event location to the camera. A schematic illustration of the algorithm is presented in Figure \ref{fig:schema_path_image}, along with an example of produced image of a cloud field.

\subsection{Insensitivity of Computing Time to the Amount of Volumetric Data}
The first radiative transfer example deals with a cloud field that is typical of today's large LES. Simulating all flow structures from turbulence at metric scales to organized convection at mesoscale (kilometric), with a potential coupling with a complex surface, is a relatively recent achievement permitted by the increase in computational power and heavy parallelization \cite{dauhut_giga-LES_2016,heinze_large-eddy_2017}. These high-resolution, large-domain simulations open new perspectives but come with limitations related to the amount of produced data. Post-treatment and analysis is getting difficult, and the outputs of such simulations are not always exploited to their full potential, at least as far as studies of cloud-radiation interactions are concerned. This is clearly one of the motivations that led us to develop radiative tools that scale with this increasing amount of data.

\begin{figure}\centering
	\includegraphics[scale=.6]{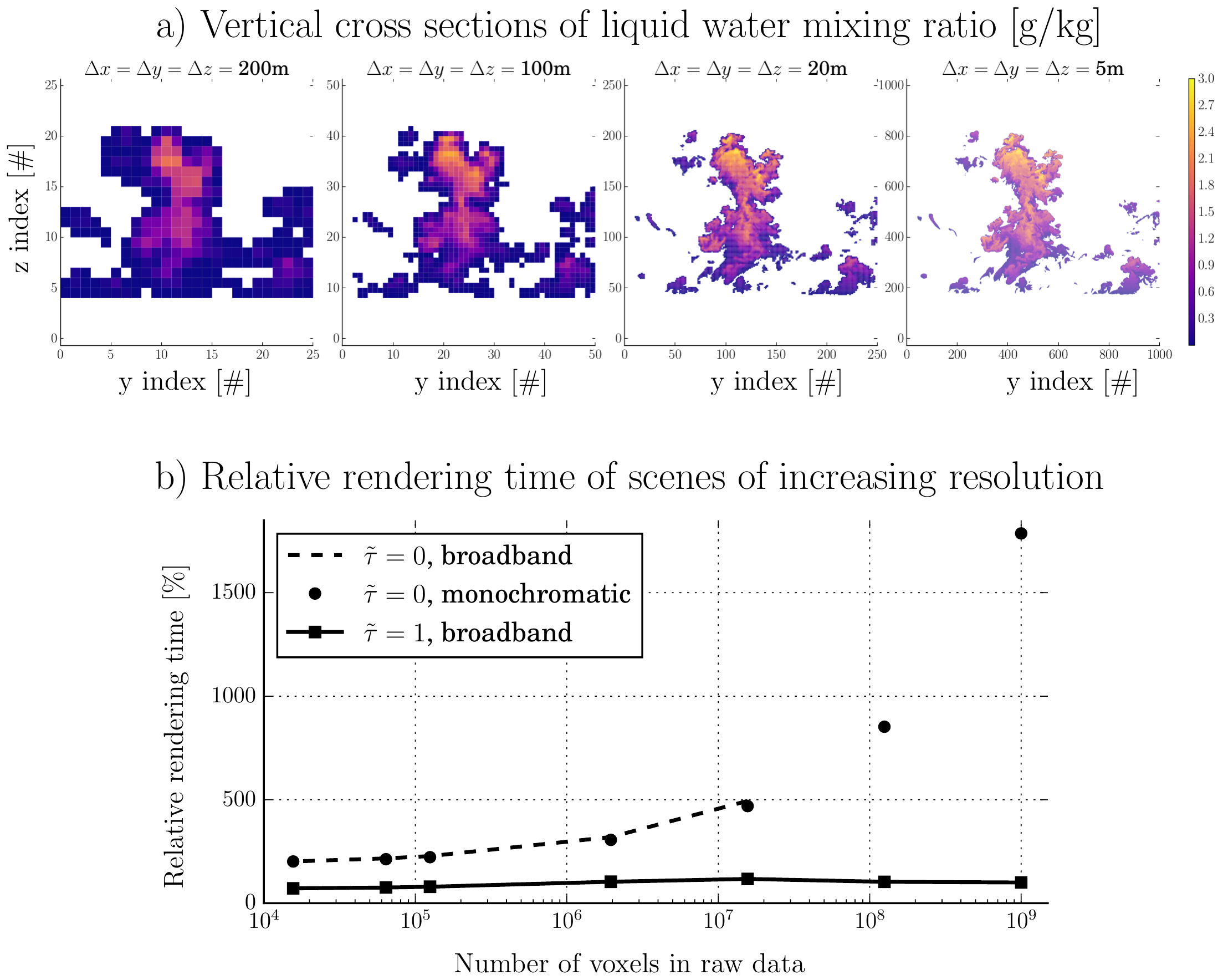}
	\caption{a) Vertical cross sections of liquid water content representing cloud fields of increasing resolution (top). b) Mean rendering time of a realisation (path) as a function of the number of cells in the volume, relatively to the mean rendering time of a path tracked in the scene using the data of highest resolution (5 m, 1000x1000x1000 cells) (bottom). Full-line results: hierarchical grids with optical depth merging criterion of 1. Dashed-line results: hierarchical grids with optical depth merging criterion of 0 (the full resolution of the original field is preserved). Without merging cells, rendering could not be achieved in the broadband configuration for scenes with resolution under 10 m: the thirty hierarchical grids (one per quadrature point) could not fit into memory. To extend the plot to 5m- and 10m-resolution fields, monochromatic computations (black dots) were performed: only one grid needs to be stored, therefore the computation becomes affordable.}
	\label{fig:insensibilite_volume}
\end{figure}

Figure \ref{fig:insensibilite_sol} already has illustrated that the computing time of a radiative calculation based on Monte Carlo techniques can be insensitive to the complexity of the surface representation. The main objective of our developments was to retrieve this same characteristic for volumes. Figure~\ref{fig:insensibilite_volume} represents the evolution of the computing time needed to render the congestus cloud scene shown in Figure~\ref{fig:schema_path_image}-b), as a function of the size of the LES grid.

Starting from the 5m-resolution output from Meso-NH shown in Figures~\ref{fig:hierarchical_structure}-a) and \ref{fig:schema_path_image}-b), the 3D fields of temperature, pressure, vapour and liquid water are artificially coarse-grained to obtain fields of lower resolutions (down to 200 m). In each voxel of the coarse resolution fields, regional averages of the high resolution fields are stored. The domain size remains constant, only the resolution and hence the number of cells in the field are changed. Illustrations of some of the resulting cloud fields are shown in Figure~\ref{fig:insensibilite_volume}-a). Since cloudy cells are averaged together with clear cells near the cloud edges, the volume of the cloud increases while the resolution decreases, but the total liquid water content is unchanged. Hierarchical grids are then built for the different cloud fields, with a criterion on the merged voxel optical depth of either: \begin{enumerate}
\item $\tilde\tau=1$: voxels are merged while the vertical optical depth of the merged region is less than 1,
\item $\tilde\tau=0$: voxels are never merged hence the hierarchical grid is at the same resolution as the original data grid. \end{enumerate}

Fields of radiances are then rendered with the same camera and sun set-up as for the image shown in Figure~\ref{fig:schema_path_image}-b). The same number of pixels and paths per pixel is used: the resolution of the radiance field is independent from the resolution of the cloud field itself. Figure \ref{fig:insensibilite_volume} shows that the rendering time for computations with merged hierarchical grids is almost constant while the rendering time for computations with unmerged hierarchical grids increases exponentially with the resolution of the field. Sensitivity of the computing time to the merging criterion $\tilde\tau$ is further investigated in the next subsection.

\subsection{Comparative Tests for Typical Boundary-layer Cloud Fields}
\label{part:images_tau}
\begin{figure}\centering
	\includegraphics[scale=1]{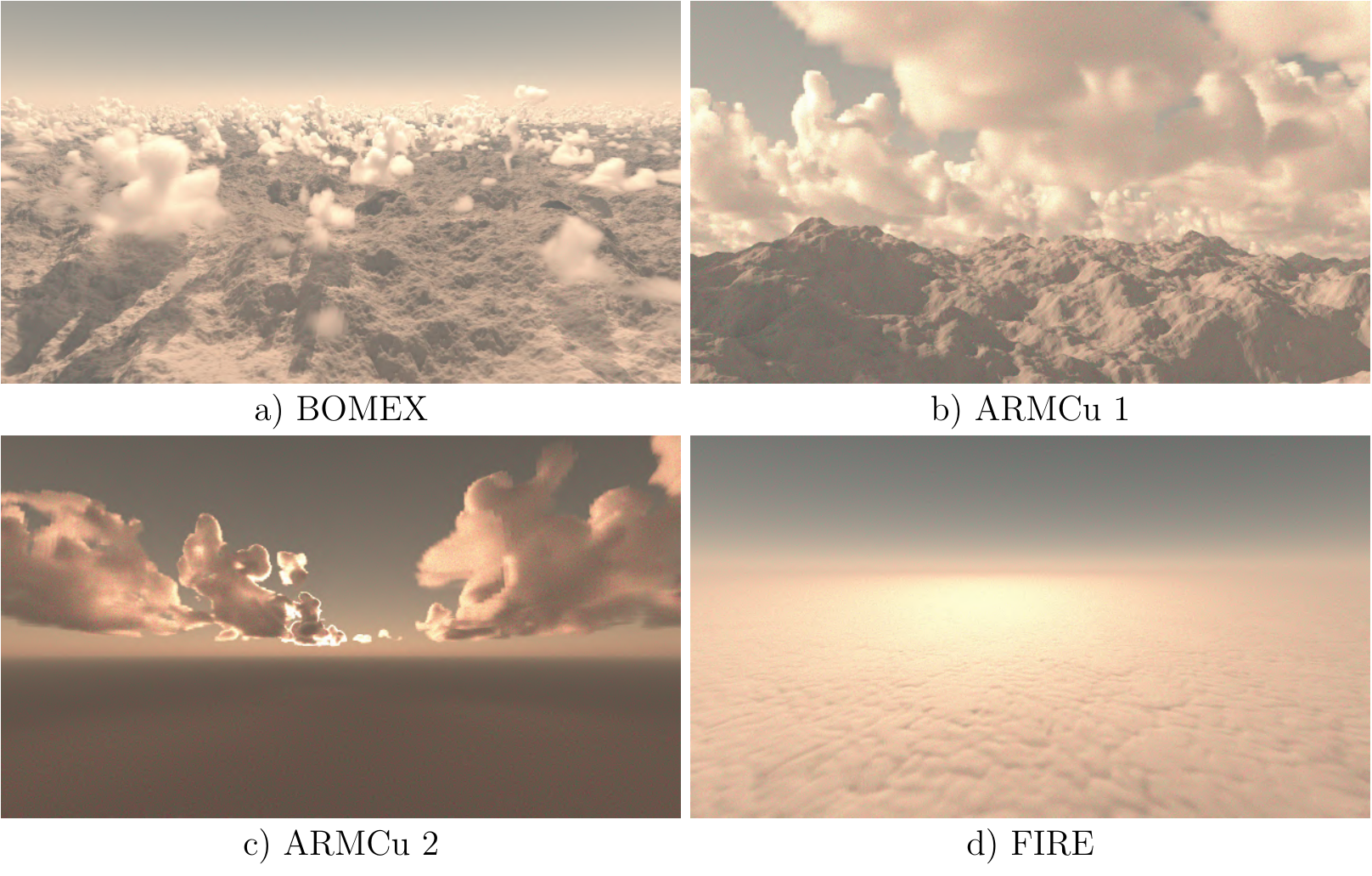}
        \caption{Rendering of LES fields from the a) BOMEX, b) and c) ARMCu and d) FIRE cases. The ground is complex in a) and b) (2x2048x2048 triangles) and plane in c) and d) (2 triangles). Camera configurations and sun positions are summarized in Table~\ref{table:scenes} of Appendix~\ref{app:lib}. They are the same as in the scenes from the starter-pack, available online. For all images, the definition is 1280x720 pixels, with 4096 samples per pixel component (and 3 components per pixel).}
	\label{fig:imgs}
\end{figure}

Next performance tests make use of idealized LES fields that are representative of the diversity of boundary-layer cloud regimes: continental cumulus clouds (ARM-Cumulus, \citeA{brown_large_2002}) run at 25 m resolution; marine, trade-winds cumulus at 25 m resolution (BOMEX, \citeA{siebesma_large_2003}); and a stratocumulus case at 50 m resolution (FIRE, \citeA{duynkerke_observations_2004}). They are less challenging than the previously studied congestus in terms of amount of data (respectively 256$\times$256$\times$160, 512$\times$512$\times$160, 250$\times$250$\times$70 grid cells), but they are typical of our practice of using high-resolution simulations to study small scale processes and support the development of parameterisations in larger-scale models. Low clouds are of particular interest since they are a frequent regime in time and space and their radiative impact is key to the energetic balance of the Earth system, and hence to the the evolution of its climate \cite{bony_marine_2005}. In the field of study of cloud-radiation interactions, each new question or observable can lead the specialist to design an entirely new Monte Carlo algorithm. It is to answer this need, with the objective of insuring that our tools be flexible enough in their use, that our developments are at the library level and not at the application level. However, the benefit of using this library in terms of acceleration should not depend on the type of cloud that is studied. Here, we show how the path-tracing library, through the rendering algorithm presented before, behaves when confronted to various liquid clouds, from thin marine cumulus to thicker stratocumulus clouds. Image of these scenes are shown in Figure \ref{fig:imgs}. The renderer is applied to the same cumulus field in Figures \ref{fig:imgs}-b) and \ref{fig:imgs}-c), but the surface is plane in \ref{fig:imgs}-c) while it represents a complex terrain in \ref{fig:imgs}-b).

For each image, Table \ref{table:times} gives the average time per realisation (path), its standard deviation (computed over all realisations), the total rendering time over 40 threads and the equivalent speed in number of realisations per second. Since we have shown that the amount and complexity of surface or volumetric data does not impact the rendering time, the only source of variation for the different cloud scenes are: \begin{enumerate}\item the camera settings: cloudy pixels take longer to render than clear-sky pixels because of the high order multiple scattering \item the clouds themselves: overcasting cloud layers or optically thick clouds take longer to render than broken or thin clouds because paths get more often trapped which increases the order of scattering and hence the length of the path.\end{enumerate} Indeed, the BOMEX field is four times larger than the ARMCu field and shows an equally complex surface, yet it is the scene that presents the shortest rendering time.

\begin{table}
        \caption{Rendering times for images of various cloud scenes.} 
	\begin{tabular}{l|cc|cc}
    \toprule
		Image & $\overline{t}$ $[\mu s]$ & $\sigma_t$ $[\mu s]$  & Total rendering time & speed $[\#path.s^{-1}]$\\
    \midrule
		Congestus 5m & 117.986 & 0.0052  & 9h22 & 335 842 \\
		BOMEX & 37.255 & 0.001 &  2h59 & 1 054 433 \\
		ARMCu 1 &  105.049 & 0.0018 &  8h22 & 375 983 \\
		ARMCu 2 & 60.425 & 0.001 &  4h59 & 631 249 \\
		FIRE  & 122.061 & 0.0016 &  10h01 & 314 049 \\
    \bottomrule
          \multicolumn{5}{l}{Images were computed with 3 (channels) x 1280x720 (pixels) x 4096 (paths)}\\
          \multicolumn{5}{l}{= 11 324 620 800 sampled paths, over 40 threads of a CPU clocked at 2.2 GHz.}\\
          \multicolumn{5}{l}{Times per realisation $\overline{t}$ and their standard deviations $\sigma_t$ are given for one thread,}\\
          \multicolumn{5}{l}{total rendering time and speed are given for parallel computation over 40 threads.}
	\end{tabular}
  \label{table:times}
\end{table}

As stated in Section~\ref{part:grilles_expectations} the acceleration potential of null-collisions used in combination with hierarchical grids depends on a compromise between the cost of traversal of the grid (increasing with the hierarchical grid resolution e.g. when fewer voxels are merged), and the cost of rejecting many null-collisions (increasing when too many voxels are merged). This ratio of costs is therefore controlled by the construction strategy of the hierarchical grid. We show how rendering time, and its partionning into crossing voxels and rejecting null-collisions, are impacted by the optical depth threshold used to merge voxels when building the hierarchical grids. 

\begin{figure}
	\centering
	\includegraphics[scale=1]{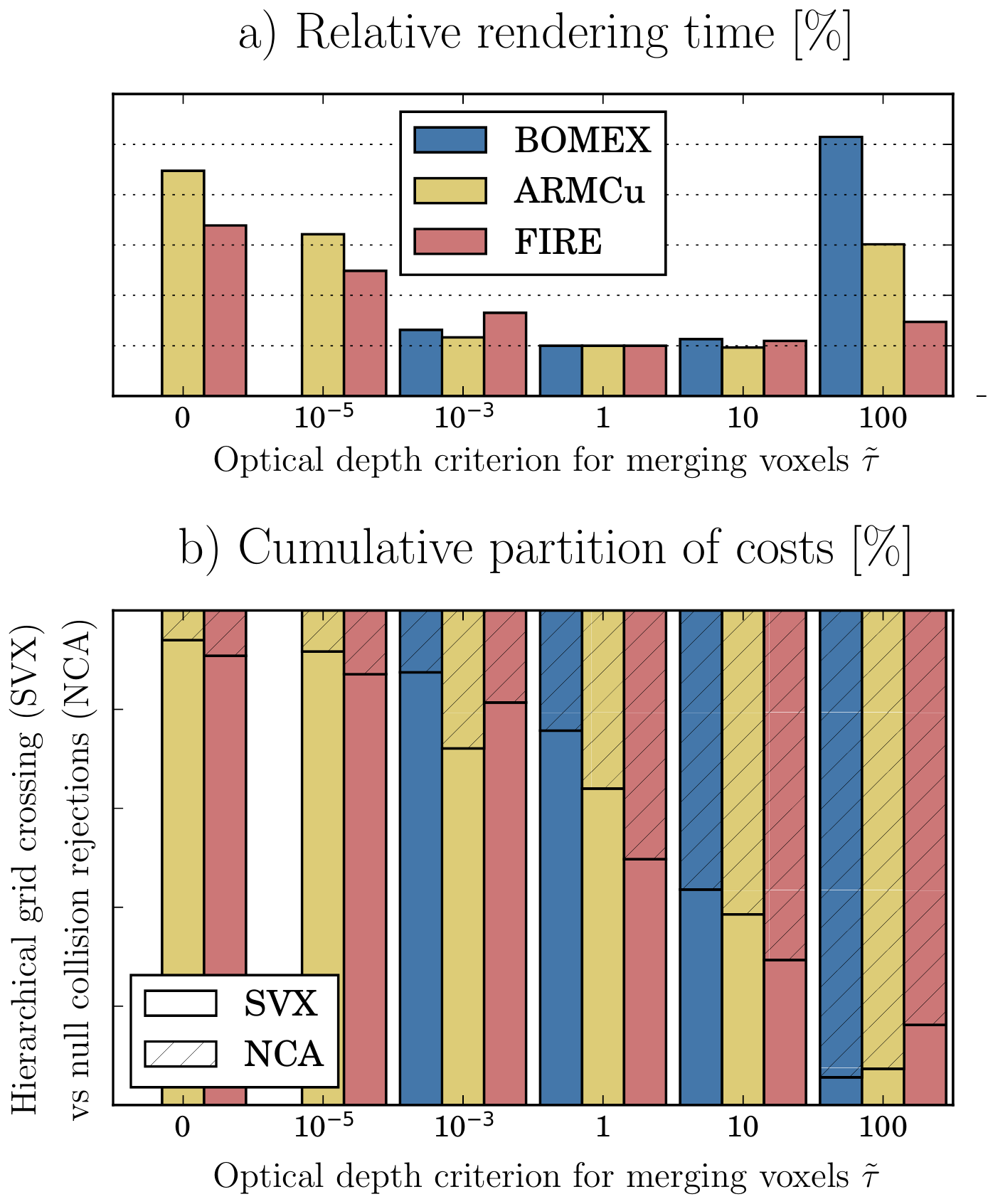}
	\caption{a) Dependence of computing time (top) and b) its partition (bottom) into i/ crossing and accessing acceleration structure voxels (SVX) vs ii/ accessing raw data and testing collision nature (NCA), to the optical depth threshold $\tilde\tau$ used as a merging criterion during hierarchical grid construction. Small values for this limit correspond to refined structures. Note that BOMEX values are missing for $\tilde\tau \leq 10^{-5}$ because the thirty hierarchical grids (one per quadrature point) did not fit into the main memory (the BOMEX fields are 4 times larger than the ARMCu fields).}
	\label{fig:perf}
\end{figure}

Figure~\ref{fig:perf}-a), shows that an optimum value for $\tilde\tau$ seems to lie between 1 and 10 for all the tested scenes. For these values, grids are such that one to ten collisions occur in average in each voxel. If, for all cloud fields, computations are faster when using an optimum hierarchical grid, computing times for larger $\tilde\tau$ show that fields with lesser volumic fractions of cloudy cells (e.g. BOMEX) benefit more from the hierarchical grids than globally cloudier fields. Computational times for smaller values of $\tilde\tau$ show, as in Figure \ref{fig:insensibilite_volume}, that fields described by larger datasets benefit more from the hierarchical grids. Looking at the partitionning into i/ crossing and accessing acceleration structure voxels (SVX) vs ii/ accessing raw data and testing collision nature (NCA), Figure~\ref{fig:perf}-b) shows that as expected, the optimum strategy for building hierarchical grid is between the limits of systematically crossing each cell of the original data, and using a fully homogeneized collision field.

\section{Other Examples of Implementation for Cloud-interactions Studies}
\label{part:perspectives_other}
As a perspective, we show here examples of Monte Carlo algorithms evaluating other metrics that we use in the context of 3D cloud-radiation studies. These additional simulation examples are only presented as perspectives and the corresponding computational performances will not be discussed because they were performed using an intermediate version of the library. Their recoding with the new version is ongoing.

\subsection{Parametric Sensitivities}
\label{subpart:sensib2}
\begin{figure}
\centering
\includegraphics[scale=.5]{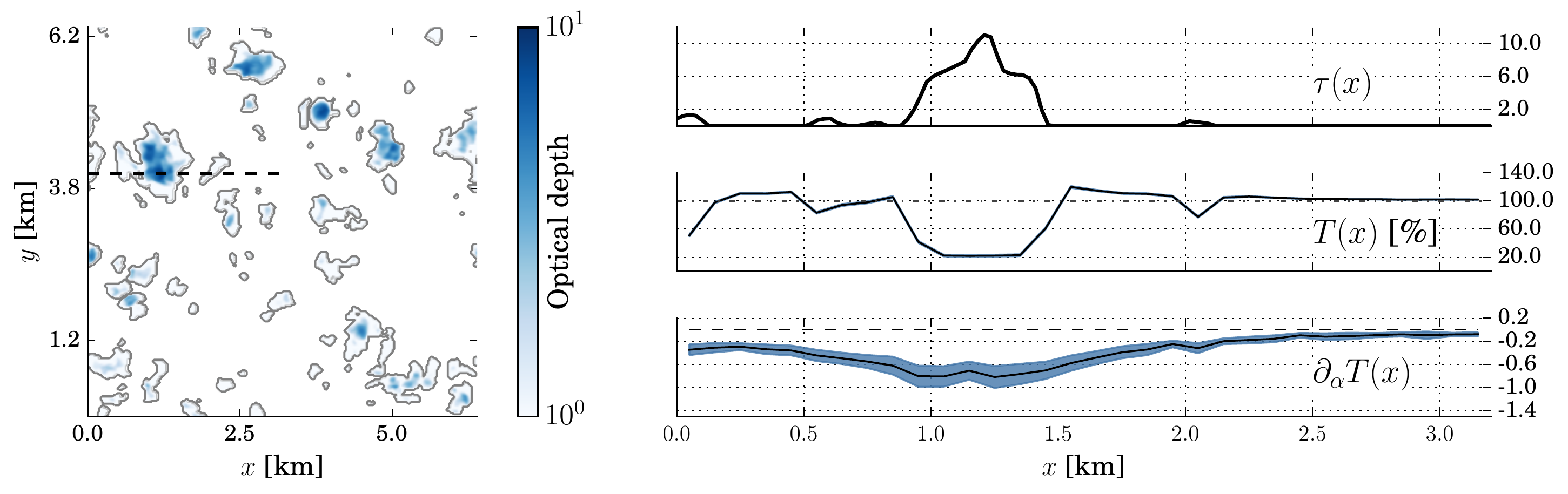}
        \caption{Left: horizontal map of the optical depth (in logarithmic scale) for a cumulus case (ARMCu 6th hour, 1330 Local Time). Right: optical depth $\tau$, transmissivity $T$ and its sensitivity to the absorption-to-extinction ratio along the dashed line shown on the left figure. Blue shaded areas represent the 3$\sigma$ confidence interval estimated by Monte Carlo. The estimation of the transmissivity and its derivative was achieved without approximation, through a unique Monte Carlo computation.}
\label{fig:sensib}
\end{figure}

The idea of evaluating sensitivities has been introduced in Section~\ref{part:introduction} and used in Section~\ref{part:grilles} to illustrate the practical meaning of filtering functions. A first example of such sensitivity simulations is implemented, starting from a backward Monte Carlo algorithm estimating the monochromatic ground flux density at a given location. Clear-sky optical depth is set to zero, only clouds interact with radiation. The sun, directed along a given direction, illuminates the TOA uniformly. Using integral developments such as those of Equation~\ref{eq:intform-sensib}, only a few extra code lines were required to implement the simultaneous computation of the sensitivity to the absorptivity (single scattering albedo) of cloud droplets, $\alpha$ (the ratio of absorption over extinction coefficients). This parameter is an output of Mie computations, that rely on hypothesis such as droplet size distribution and purity, and is typical of the radiative transfer uncertainties associated to cloud microphysics modeling. Results of a simulation in a cumulus cloud field with the sun at the zenith are displayed in Figure \ref{fig:sensib}.

Evidences of 3D effects appear in Figure \ref{fig:sensib}: transmissivity in clear sky is greater than 1 near the clouds due to sideways leakage of photons through cloud edges. The transmissivity sensitivity to the absorption ratio is negative since more absorption compensated by less scattering yields less total transmissivity, with a maximum under the cloud and slow return to zero elsewhere. The fact that this sensitivity is non-zero under clear sky conditions is another evidence of the remote horizontal influence of clouds on local radiation.

Since algorithms are in general optimized to produce a low variance on the quantity estimate, there is no guarantee that the variance of the derivative estimate will also be low. We see in Figure \ref{fig:sensib} that the three standard deviations interval ($\pm 3\sigma$, represented as a shaded area around the curves) is much more important around the sensitivity than around the quantity itself. This is due to known difficulties associated to the computation of sensitivities in highly scattering media. Investigations to efficiently reduce this variance without losing convergence on the quantity itself is currently undergoing.

\subsection{Evaluation of a Large-scale Radiative Transfer Parametrization}
The developed tools have also been used to compare reference Monte Carlo results to computations from the radiation scheme ecRad \cite{hogan_flexible_2018}. Possible solver choices implemented in ecRad include Tripleclouds, a 1D two-stream solver that represents subgrid horizontal variability of the medium by defining three regions in each layer \cite{shonk_tripleclouds_2008} and the SPARTACUS solver \cite{schafer_representing_2016,hogan_representing_2016} based on Tripleclouds but that additionally represents the effect of horizontal transport on the vertical fluxes. Here, MC calculations in a cumulus cloud field that is used as a reference to evaluate ecRad and its parametrization of 3D effects in the estimation of the direct-to-total fluxes ratio at the surface are presented. Relevant cloud parameters such as overlap and cloud scale are diagnosed in the LES field and provided to ecRad. 

In the broadband solar forward MC, direct and diffuse fluxes are horizontally integrated: paths contribute to the same estimate independently of their horizontal location when they hit the surface. To allow comparison, wavelengths are sampled according to the Rapid Radiative Transfer Model for GCMs (RRTMG, \citeA{mlawer_radiative_1997, iacono_radiative_2008}) k-distribution model, in the solar interval ([820-50000] cm$^{-1}$). Input gas profiles are taken from the I3RC cumulus case file provided with the ecRad package. Only vertical variations of gas absorption coefficients are considered. A path contributes to the diffuse flux if it has been scattered at least once. A difficulty in comparing MC to ecRad resides in the fact that, in solvers based on the two-stream model such as Tripleclouds and SPARTACUS, the partition between direct and diffuse fluxes is often biased. Indeed, using only two slantwise directions to propagate the diffuse fluxes leads to an underestimation of transmissivity due to the fact that, in reality, clouds scatter a large amount of radiation in a very small solid angle around the forward direction.

\begin{figure}
\centering
	\includegraphics[scale=.5]{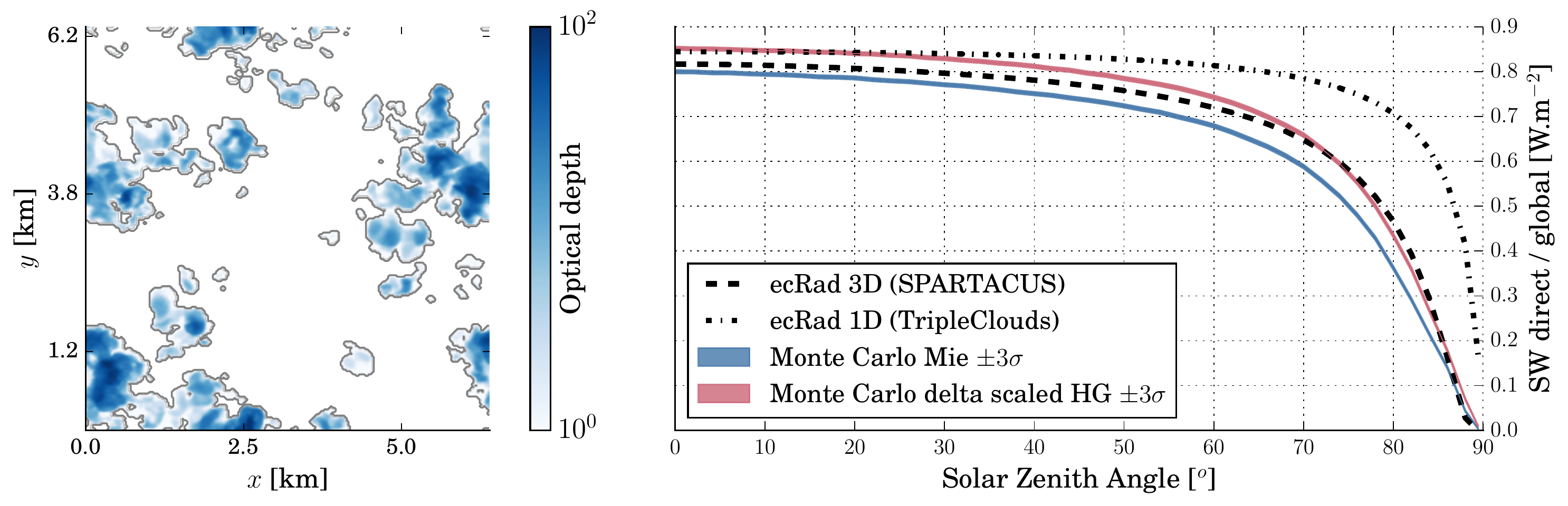}
        \caption{Left: horizontal map of the optical depth (in logarithmic scale) for a cumulus case (ARMCu 8th hour, 1530 Local Time). Right: Monte Carlo vs. ecRad computations of surface horizontally averaged direct-to-total broadband fluxes ratio, as a function of solar zenith angle. Results from two ecRad simulations with different solvers (Tripleclouds and SPARTACUS) are plotted to evidence the impact of 3D effects on the partition of surfaces fluxes. Results from two Monte Carlo simulations with different phase functions (Mie or delta-Eddington scaled Henyey-Greenstein) are plotted to assess the bias related to the delta-Eddington scaling approximation.}
\label{fig:ratio}
\end{figure}

The delta-Eddington scaling technique \cite{joseph_delta-eddington_1976} is generally applied to correct for the too reflective clouds: the optical depth and asymmetry parameter are reduced to artificially avoid the scattering of some of the forward scattered photons, leading to a correct estimation of the total ground flux, but to an overestimation of the direct component. In addition to the exact MC computation using the true Mie phase function, a MC simulation using a delta-Eddington scaled Henyey-Greenstein phase function is performed to assess the bias related to the delta-Eddington scaling approximation. The cloud field optical depth and results are shown in Figure \ref{fig:ratio}.

The evolution of the direct-to-total fluxes ratio at the surface is plotted as a function of the solar zenith angle. Since the effective cover increases when the sun is lower in the sky, more of the direct beam is intercepted by clouds through their edges. Without 3D effects, ecRad fails to represent the reduction of the direct contribution with respect to the solar zenith angle. Good agreement is found between MC and ecRad when these 3D effects are represented i.e. when the SPARTACUS solver is used, effectively reducing the amount of direct radiation at large solar zenith angles. As expected since the optical depth has been reduced in the approximation, using a delta-Eddington scaled HG phase function instead of the true Mie phase function yields an overestimated direct flux at the surface.

\section{State of the Library and Current Limitations}
\label{part:perspectives_limitations}

Through the developments presented in this paper, our contribution to the field of atmospheric radiative transfer is as follows:
\begin{enumerate}
	\item We revisit null-collision (maxium cross section) algorithms with recursive, hierarchical grids (octrees) to accelerate ray casting in volumes: this makes the computing time independent of the data amount.
	\item We value the practice of writing the integral formulation that is equivalent to the Monte Carlo algorithm. In its integral form, the null-collision method appears as more than a rejection method but as a way to bypass Beer's law nonlinearity. Simultaneous evaluation of sensitivities is given as an example of algorithm derived from integral reformulation.
	\item We provide low-level libraries and not one code to allow the required flexibility in the implementation. The filtering function abstraction creates a true separation between the algorithm and the ray casting procedure.
\end{enumerate}

We freely distribute our library online, together with atmospheric data and a rendering code that produces synthetic images of cloud fields. It is coded in C, for CPU technology. Part of the library is based on Embree. The low level modules (detailed in Appendix \ref{app:lib}) are elementary bricks that can be used for a large range of applications. They implement well separated concepts and are easily maintained. The modules that are related to the atmosphere, e.g. \textit{sky} that builds the hierarchical grids from a combination of 1D clear profiles, 3D cloud fields and spectral properties in the form of k-distributions, will evolve as needs for improvement arise. For instance, in the current version of \textit{sky}, one hierarchical grid is constructed per spectral quadrature point. This choice should be revisited and the cost of constructing the octrees should be weighed against the cost of having only one grid for the whole spectrum and hence a suboptimal volumetric structure. Strategies for combining varying optical properties across the spectrum into one grid should be investigated.

Currently, the various octrees (thirty when solving radiation in the visible solar spectrum) are distributed to the various active threads that construct them in parallel. Although this was not implemented yet, parallelized construction of one octree should be trivial since each octree can be seen as a parent-octree partitionning independent children-octrees hence an arbitrarily large number of children-octrees can be constructed simultaneously by various threads. 

To be able to handle large datasets that do not fit into main memory, as for instance full 3D atmospheres with multiple cloud layers, and energetic systems such as solar plants on the ground, one has to adopt the \textit{out-of-core} paradigm for the whole library, i.e. all the data are stored on disk and (un)loaded on demand. As described in Section~\ref{part:liblib}, both the raw data and the acceleration grids were thought with this perspective: cloud fields can already be dynamically loaded from disk, and acceleration grids are carefully structured with this objective in mind.  However, this will only be efficient if the algorithms that require data access are designed according to the out-of-core nature of the data. For instance, the strategy implemented in Hyperion, Disney's out-of-core renderer \cite{burley_design_2018}, consists in tracking rays in bundles, with many rays visiting the same regions and hence making intensive use of the loaded data before unloading it when memory space runs out.

Regarding the physics, algorithmic developments could be undertaken to improve the convergence of the estimators described in our examples. We do not expect any technical difficulty in implementing existing or new solutions to e.g. the convergence issues related to the peaked Mie phase function in backward solar algorithms \cite{iwabuchi_fast_2009,buras_efficient_2011}. However, the question of the sensitivities slow convergence rate deserves further investigation. Estimating other types of sensitivities could also be explored, for instance geometric sensitivities instead of parametric ones: the uncertain parameter is no longer under the integral but is part of the integration domain, for example the position, inclination or field-of-view of the camera. Computations in the longwave part of the spectrum should be straightforward, although the strategy for constructing the hierarchical grids might need some investigation, since the heterogeneous temperature field is added to the optical heterogeneities. Even in the shortwave, more work is needed to deepen our understanding of which strategy is most appropriate for building the grids, depending on the cloud field, its spectral properties, and the algorithm. The treatment of ice crystals, aerosols or varying liquid droplet size distribution would require extending the library to load additional 3D fields, and the data to add a dimension to the Mie look up table describing the cloud optical properties. We do not expect any technical difficulty here either, but this has yet to be developed. Indeed, our focus was until now on the ray-casting procedure, but further developments should yield a more comprehensive toolbox where more complex atmospheric fields can be handled.

\appendix
\section*{Appendix}
\label{part:appendix}

\section{Brief History of Path Tracing in Surfaces and Volumes}
\label{app:history}
The content of this appendix is not a rigorous review. Our understanding of the history of \textit{path tracing} inside scenes involving large geometric models of opaque surfaces, is briefly summarized, with a specific attention to the computer science literature devoted to physically-based rendering, which indeed addresses the very same radiative transfer equation as ours (\ref{part:MC_surf}). Recent developments made in the handling of complex volumes by both this community and the engineering physics community (for infrared heat transfer and combustion studies) are then listed in \ref{part:MC_vol}. Based on our understanding of this literature, a non comprehensive compararative table of the state of the art of both communities: computer graphics and atmospheric radiative transfer, is presented in (\ref{part:MC_comp}).

\subsection{Path Tracing and Complex Surfaces}
\label{part:MC_surf}
Image synthesis is the science that aims to numerically produce images from descriptions of scenes. It was born in the 1970s when computer graphics started to expand. At first, the focus was on surface rendering, often assuming the scene objects were surrounded by vacuum. Among the diverse existing techniques, we mention here only a few ones that gradually led to the use of Monte Carlo based path-tracing methods to render 3D scenes. Methods that were dominant in practice (e.g. micropolygon rendering or rasterization) are missing from this text and we refer the interested reader to more complete presentations of the field's history, e.g. in the Section 1.7 of \citeA{pharr_physically_2018}.

The initial concern was to determine which objects in a scene were visible from a given point of view. \citeA{appel_some_1968} first introduced the \textit{ray casting} method as a general way to solve the \textit{hidden surface problem}, by casting rays from the observer to the scene objects and detecting intersections. This opened a whole field of investigation dedicated to optimizing ray casting, e.g. through efficient intersection tests between rays and large numbers of primary shapes (see \citeA{wald_interactive_2001,wald_realtime_2004,wald_embree_2014} and references therein).

Once the visible surfaces were found, the next question was to determine how these surfaces were illuminated by the sources and the other surfaces, which was referred to as the \textit{global illumination problem}. \citeA{whitted_improved_1980} first used recursive ray casting in the \textit{ray tracing} method, which includes random sampling around optical directions to correct the unrealistically sharp gradients of intensity due to perfectly specular reflections. \citeA{cook_distributed_1984} then generalized the randomly perturbed ray-tracing approach to multi-variate perturbations in the \textit{distributed ray tracing} method. This was the first algorithm able to render all the major realistic visual effects in a unified, coherent way.

A couple of years later, \citeA{kajiya_rendering_1986} developed the formal framework of the \textit{rendering equation} (the integral formulation of the radiative transfer equation in vacuum, focused on light-surface interactions). His \textit{path tracing} model was the first unbiased scene renderer to be based on MC ray casting. While revisiting this proposition, \citeA{arvo_particle_1990} found inspiration in the experienced community of particle transport sciences, where MC methods were already commonly used and studied. They introduced variance reduction techniques to the image rendering community. 

Another important step toward efficiency was Veach's pioneering thesis \cite{veach_robust_1998}. From his mathematical background, he introduced a new paradigm in which radiative quantities were formally expressed as integrals over a \textit{path space}, decoupling the formulation from the underlying physics: the formulations were no longer \textit{analog} (i.e., based on intuitive pictures of the stochastic physics of particle transport). This allowed him to explore sampling strategies in full generality and to then apply them to path tracing, giving birth to several low-variance algorithms such as the \textit{Bidirectional Path Tracing} \cite{veach_bidirectional_1995} or the \textit{Metropolis Light Transport} \cite{veach_metropolis_1997}

It is only from the years 2000s, with the increase of computing power, that MC physically-based path-tracing techniques were considered viable tools beyond research, for production purposes. They were favored because \begin{enumerate} \item it was eventually perceived that MC methods allow independence between the rendering algorithm and the description of the scene (i.e. the number and properties of the surfaces to render), thus providing the artists with unprecedented freedom, \item they allow a unified, physical representation of the interaction of light with surfaces, relieving the artists from the need to arbitrarily modify the surface properties in order to achieve a specific effect, since they could now rely on the physics and \item improvement of filtering methods have allowed cheap image denoising, thus bypassing the need for more expensive, well-converged MC simulations. \end{enumerate}

\subsection{Path Tracing and Complex Volumes} 
\label{part:MC_vol}
A major difficulty in MC methods is the treatment of complex heterogeneities in volumes, e.g. cloudy atmospheres. For decades, the computer graphics handled the question of volumes as have many other MC scientists; their expertise in designing performant ray-casting tools had found its limit in dealing with volume complexity. In Section \ref{part:grilles}, it is claimed that the issue resides in the nonlinearity of Beer's law of extinction: the expectation of a nonlinear function of an expectation can no longer be seen as one expectation only. It is then stated that the method of null collisions can be seen as a way to bypass Beer's nonlinearity.

In neutron transport, this method was first described by \citeA{woodcock_techniques_1965} under the name Woodcock tracking. In plasma simulations it first appeared in \citeA{skullerud_stochastic_1968}. Soon after, \citeA{coleman_mathematical_1968} gave a mathematical justification for this method, demonstrating its exactness. In the atmosphere, it was first proposed by \citeA{marchuk_monte_1980} and called the \textit{Maximum cross section}. \citeA{koura_null_1986} developed it for rarefied gas under the name null-collisions. Computer graphics have also used it as Woodcock tracking, for the first time in \citeA{raab_unbiased_2006}. 

Only with \citeA{galtier_integral_2013} seminal paper did it become clear that null-collision methods allowed a reformulation of the integral solution to the radiative transfer equation in which the difficulties related to the nonlinearity of Beer's law disappear. In this paper, null-collision algorithms (NCA) are written as integral formulations, and it is shown that the null-collision methods can be used in a more flexible way, including with negative null-collision extinction coefficients. It is stated that the data--algorithm independence, also strongly highlighted by \citeA{eymet_null-collision_2013}, is not a consequence of introducing null-collisions, but rather a consequence of the underlying integral reformulation.

This explicit framework opened doors to new families of MC algorithms, with potential for solving various problems that were before then considered impossible: nonlinear  models \cite{dauchet_addressing_2018}, coupled radiation-convection-conduction in a single MC algorithm \cite{fournier_radiative_2016}, energetic state transitions sampled from spectroscopy instead of approximate spectral models \cite{galtier_radiative_2016}, symbolic Monte Carlo to scattering media \cite{galtier_symbolic_2017} etc. Some of these methods are transposable to atmospheric radiative transfer with large benefits for our community, e.g. conducto-radiative MC models to investigate atmosphere--cities interactions, or line-sampling methods for benchmark spectral integration, to develop, tune and test spectral models. During the past couple of years, the computer graphics community has been similarly impacted by this new paradigm. \citeA{kutz_spectral_2017} show how integral formulations of NCA can be used to derive more efficient free-path sampling techniques. \citeA{novak_monte_2018} give a good review of the different free-path sampling methods, with a focus on NCA and their newly perceived interest: acceleration structures that were already used for surfaces could now be used for volumes. 

\subsection{Comparison of the Computer Graphics and Atmospheric Science Literatures}
\label{part:MC_comp}
A non comprehensive summary of contributions from the computer graphics and atmospheric radiation is presented in Table~\ref{table:comparison}. Only the techniques related to the library are cited. Other techniques such as variance reduction methods are mentioned in the text but do not appear in Table~\ref{table:comparison}.
\begin{table}
	\caption{Summary of techniques used in computer graphics that we make available to the atmospheric community through our library}
	\begin{tabular}{lll} \toprule
		\textbf{Method}& \textbf{Computer graphics} & \textbf{Atmospheric radiation}\\ \midrule
		Null-collision algorithms & Woodock tracking & Maximum cross section \\
		&  \citeA{raab_unbiased_2006} & \citeA{marchuk_monte_1980} \\\hline
		Acceleration for surfaces  & Bounding Volume Hierarchy & No standard \\
		& \citeA{wald_embree_2014} & \citeA{mayer_validating_2010}\\ && \citeA{iwabuchi_modeling_2006}\\\hline
		Acceleration for volumes  & Octrees & No standard\\
		& \citeA{burley_design_2018} & \citeA{iwabuchi_multispectral_2017}\\\hline
		Memory management         & Out-of-core & - \\
		& \citeA{baert_out-of-core_2013}& \\ 
    \bottomrule
\end{tabular}
	\label{table:comparison}
\end{table}

\section{Physical and Optical Properties of the Cloudy Atmosphere}
\label{app:data}
As mentioned in the text, our Monte Carlo codes handle liquid clouds and atmospheric gas, which production in terms of contents and optical properties we describe below. This data are provided with the library since it was used in all the tests that were performed up to now. The nature of the data has conditioned choices, mainly proper to the applications, that we expose below. The only particularity in the implementation of the low-level libraries themselves is that, due to the fact that we provide 3D cloud fields and 1D gas profiles, the library can manage 3D and 1D data. The sky module combines the 3D and 1D data wherever the domains intersect each other, and uses low level procedures to build the hierarchical structures.

\subsection{Physical Properties of the Atmosphere}
\label{part:clear}
\subsubsection{Clear-sky} 
The clear-sky atmospheric column is described from ground to space by vertical profiles of temperature, pressure, water vapour mixing ratio, and a mix of other gases ($CO_2$, $CH4$, $N_2O$, $CFC1$, $CFC2$, $O_2$, $O_3$). The I3RC cumulus case profiles provided with the ecRad package (the radiative transfer model developed at the ECMWF \cite{hogan_flexible_2018}) are used.
\subsubsection{Clouds}
The realistic 3D cloud fields, are produced by the Meso-NH model \cite{lafore_meso-NH_1997,lac_overview_2018} used in a Large-Eddy Simulation (LES) mode, at resolutions lying between 5 and 50m. The subgrid microphysics is a bulk, one-moment scheme (ICE3, \citeA{caniaux_numerical_2013}). No subgrid cloud scheme is used, i.e. the cells are assumed to be homogeneously filled with condensate water when saturation is reached. The 3D turbulent scheme \cite{cuxart_turbulence_2000} is closed with a mixing length based on \citeA{deardorff_stratocumulus_1980}. The model outputs 3D fields of liquid and vapour water mixing ratio, potential temperature and pressure.

\subsection{Optical Properties of Gas and Clouds}
\subsubsection{Gas Molecules}
The radiative properties of the atmospheric column are computed according to the Rapid Radiative Transfer Model for GCMs (RRTMG, \citeA{mlawer_radiative_1997, iacono_radiative_2008}). We access them via the ecRad software, that we use as a front-end for production of the RRTMG k-distributions profiles for 16 spectral intervals in the longwave (LW) region ([10-3500] cm$^{-1}$) and 14 spectral intervals in the shortwave (SW) ([820-50000] cm$^{-1}$). Each quadrature point, in each spectral interval, is provided with a quadrature weight that is used by our algorithms as a probability for the sampling of absorption coefficient values that are then practically used as if radiative transfer was monochromatic. The only subtility is related to the variability of the water vapor concentration in the 3D LES domain. The water vapor mixing ratio, temperature and pressure are provided in each cell of the 3D domain. The impact of the horizontal variations of temperature and pressure on the absorption being negligible besides the impact of vapour variations (in solar computations), we only consider the vertical profiles of horizontally-averaged temperature and pressure fields to deduce the absorption coefficient profiles. For water vapor, we use the fact that the absorption coefficients of the gas mixture are roughly linear (in log/log space) with $x_{H_{2}O}$, the water vapor molar fraction, except for very small and very high values of $x_{H_{2}O}$. The ecRad software is therefore used in a preliminary step to compute and tabulate absorption and scattering coefficients for the 1D atmosphere, for each LW and SW interval and each quadrature point in each atmospheric layer, for a given discretized range of the water vapor molar fraction $x_{H_{2}O}$. The resulting look-up table is then used within the MC algorithm to rapidly retrieve the local k-values. Details describing the model and the interpolation procedure are given in the Supporting Information, along with a plot of the relative error on LW net fluxes. The maximum relative error between two profiles computed analytically from RRTM-G vs. interpolated absorption coefficients is around 1.2\%. This is around half the maximum relative error found between profiles computed by ecRad vs. analytically, from RRTM-G data (2.6\%). 

\subsubsection{Cloud Droplets}
\label{part:cloud}
The method developed by \citeA{mishchenko_scattering_2002}, implemented in Fortran as in \citeA{mishchenko_bidirectional_1999}, is used to solve far-field light scattering by spherical particles using the Lorenz-Mie theory. The main hypothesis are that the droplets are homogeneous and polarization is ignored. As with ecRad for gaseous absorption, this code is used externally to compute the single scattering albedo, the extinction coefficient (along with scattering and absorption coefficients), the asymmetry parameter and the phase function, all of these properties being averaged over the size distribution. We also compute the cumulative phase function and its inverse to allow efficient sampling of scattering directions. The MC algorithm accesses these data via look-up tables and performs spectral averaging over the narrow bands used in the k-distribution described above. This way, the Mie data are uncorrelated from the gas spectral data and the same look up table can be used with various spectral models. The particular table that is used for the simulations of Section \ref{part:images} is available as a NetCDF file in the starter-pack (\url{https://www.meso-star.com/projects/high-tune/starter-pack.html}). The size distribution is lognormal with an effective radius of 10~$\mu$m and a standard deviation of 1~$\mu$m.

\subsection{Implementation Choices Related to the Data Original Structure}
Volume inputs for our rendering application are hence i/ 3D fields of liquid and water vapour content, temperature and pressure, ii/ optical properties of liquid water droplets under the assumption of a constant size distribution in the whole domain and iii/ optical properties of the gas mixture in the form of absorption coefficient values, tabulated along height, spectral and water vapour contents dimensions. These data will be loaded through the htcp, htmie and htgop libraries respectively. To interface data from another format, the user can either convert the data to our input format or implement equivalent libraries to handle different input formats. 

Once the data are loaded, before using the library to build the acceleration structures, the three inputs are combined into intermediate grids. Since we will build one structure per quadrature point (an arbitrary choice that should be questionned and improved in further developments), we also build one intermediate grid per quadrature point. In this procedure, the intermediate data are never entirely constructed and the raw data are never entirely stored into the main memory: they remain \textit{out-of-core}, i.e. written on the disk and loaded and unloaded whenever needed. 

However, non negligible computational time is needed to construct these intermediate data: chunks of the 3D fields are loaded and unloaded, and tested against intersection with the 1D profiles of gas optical properties, that is interpolated to construct an intermediate 3D data grid containing local absorption and scattering coefficients of cloud and gas in each cell, for each quadrature point. By default, the size of this intermediate grid is $(2^n)^3$ where $2^n$ is the closest power of two above the raw 3D field size in its largest dimension. Since the accelerating structures are then built from these intermediate data by merging groups of cells, the highest resolution of the resulting accelerating structure is the resolution of the raw 3D data. Other strategies could be deployed e.g. the intermediate data could have a fixed size that would be independent from the raw 3D data, yielding slightly different acceleration structures. If it takes long to construct these intermediate data in our specific application, the construction of the octree itself is almost instantaneous.

\section{Description of the Set of Libraries}
\label{app:lib} 
The modules are briefly presented in Table \ref{tab:modules} and divided into three groups:
\begin{enumerate}
        \item low-level modules (random sampling, surface and volume data structuring and ray-tracing, scattering), implemented as libraries, forming the generic development environment, availabe at \url{https://gitlab.com/meso-star/star-engine/}. They implement true abstractions of Monte Carlo concepts that can be used regardless of the scientific field of application, but mastering their use requires some time and investment due to the level of abstraction they represent;
	\item data-oriented modules (3D atmospheric fields, cloud and gas optical properties data), also implemented as libraries although not directly available in the development environment since already oriented toward atmospheric applications. Using these modules as they are would require the user to produce data in the same format as ours. Another possibility is to code new (but similar) data-oriented modules that would match a new input data format and output the same objects as here in order to insure compatibility with higher-level modules;
        \item application-oriented modules (sky, ground, camera and sun), not implemented as libraries, developed in the context of the renderer application. They can be used for other projects implementing atmospheric radiative transfer models, in particular the sky module implements the construction of the hierarchical structures for the volume data that was loaded using the data-oriented modules.
\end{enumerate}

On the top of these modules, an application was developed (htrdr) that makes use of the different modules to implement a Monte Carlo algorithm. Typical functions associated to the different modules are cited as illustrations in Table~\ref{tab:modules}. The sources can be downloaded online (\url{https://www.meso-star.com/projects/high-tune/high-tune.html}) and user-guides are provided on the website. A starter-pack is also provided with the data and scripts necessary to reproduce the examples of Section \ref{part:images}. The set-up of the scenes are summarized in Table~\ref{table:scenes}. However, the most useful user-guide for the interested reader is the commented code that implements the renderer, using the various functions of Table \ref{tab:modules}. Indeed, this code was in part developed to illustrate the use of the different libraries and modules, to serve as a basis for futher developments, or as an example to implement new algorithms. 

\begin{table}\centering
        \caption{Summary of scenes set-up of images shown in the paper.}
  \label{table:scenes}
  \begin{tabular}{lccrrrrrrc}
  \toprule
  \textbf{Scene} & \multicolumn{2}{c}{\textbf{Sun}} & \multicolumn{7}{c}{\textbf{Camera}} \\
          & Zenith & Azimuth & \multicolumn{3}{c}{Position [km]} & \multicolumn{3}{c}{Target [km]} & FOV  \\
          & $\theta$ [$^o$] & $\phi$ [$^o$] & \multicolumn{1}{c}{X} & \multicolumn{1}{c}{Y} & \multicolumn{1}{c}{Z} &\multicolumn{1}{c}{X} & \multicolumn{1}{c}{Y} & \multicolumn{1}{c}{Z} &  [$^o$]   \\
  \midrule
  Congestus 5m & 25 & 230 & -2.89 & 1.98 & 2.53 & 7.90  & 2.14 & 1.16 & 70 \\
  BOMEX   & 40     & 0    & 2.22 & 3.68   & 1.49  & 8.21  & 4.47 & -0.39 & 70  \\
  ARMCu 1 & 60     & 225  & 10.24 & 0.61  & 0.42  & -2.98 & 6.83 & 0.84  & 30 \\
  ARMCu 2 & 85     & 130  & 4.66 & 0.97   & 0.83  & 0.45  & 7.05 & 1.58  & 70 \\
  FIRE    & 65     & 340  & -3.06& 11.70  & 3.80  & 10.86 & 3.68 & 0.47  & 70 \\
  \bottomrule
          \multicolumn{10}{l}{All images shown are constituted of 1280$\times$720 pixels and rendered using 4096 paths per pixel }\\
          \multicolumn{10}{l}{component, with 3 components per pixel. All scenes use the same Mie and clear-sky data.}\\
          \multicolumn{10}{l}{The sun azimuth angle origin is at $X>0$, $Y=0$ (to the East) and oriented to the North.}\\
          \multicolumn{10}{l}{FOV is for Field Of View. Position and target point values were rounded for readability. } \\
          \multicolumn{10}{l}{The data and files describing the scenes are distributed in the starter-pack, available online.}
\end{tabular}\end{table}

To test these tools in the context of multiple scattering, we implemented several benchmark experiments and compared our calculations against published results, e.g. Table 1 of \citeA{galtier_integral_2013}, or against the solution of the well-validated 3DMCPOL \cite{cornet_three-dimensional_2010} on the IPRT cubic cloud case \cite{emde_iprt_2018} (see Supporting Information). Agreement was found within the MC statistical uncertainty, thus validating our implementations.

\begin{sidewaystable} 
        \caption{Open-source Monte Carlo modules and examples of functions.}
        \label{tab:modules}
                \resizebox{.93\textwidth}{!}{%
		\begin{tabularx}{\textwidth}{l >{\hsize=0.4\hsize\RaggedLeft} X >{\hsize=0.22\hsize\RaggedLeft} X}
		\large \textbf{Module name} & \large \textbf{Description} & \large \textbf{Example of functions}\\
                \hline
                \textbf{Low-level} &&\\ \hline 
                Star-SamPle (ssp) & Generate reproducible sequences of pseudo-random numbers (compatible with parallelization), sample and evaluate various probability density functions. & ssp\_rng\_canonical; ssp\_ran\_exp\_pdf; ssp\_ran\_hemisphere\_cos; \\&&\\
                Star-3D (s3d) & Define shapes, attach them to a scene, trace rays in the scene, filter hits. & s3d\_scene\_create; s3d\_scene\_view\_trace\_ray; s3d\_hit\_filter\_function\_T;\\&&\\
                Star-VoXel (svx) & Define voxels, partition them into a hierarchical structure (tree), trace rays in the tree, filter hits. & svx\_octree\_create; svx\_tree\_trace\_ray; svx\_hit\_filter\_T;\\&&\\
                Star-ScatteringFunctions (ssf) & Setup, sample and evaluate scattering functions for surface and volume. & ssf\_specular\_reflection\_setup; ssf\_phase\_sample; ssf\_fresnel\_eval;\\
                \hline
                \textbf{Data-oriented} &&\\ \hline
                High-Tune: Cloud Properties (htcp) & Describe 4D atmospheric fields. & les2htcp (bin)\\&&\\
                High-Tune: Mie (htmie) & Describe the optical properties of water droplets. & htmie\_fetch\_xsection\_scattering; htmie\_compute\_xsection\_absorption\_average;\\&&\\
                High-Tune: Gas Optical Properties (htgop) & Describe the optical properties of atmospheric gas mixture. & htgop\_get\_sw\_spectral\_interval; htgop\_layer\_lw\_spectral\_interval\_tab\_fetch\_ka;\\
                \hline
                \textbf{Application-oriented} &&\\ \hline
                htrdr\_sky & Build acceleration grid for the atmospheric volume data (3D clouds embedded in 1D gas) in the context of null-collision algorithms, trace rays in the atmospheric volume, access null-collision and raw data. & htrdr\_sky\_create; htrdr\_sky\_fetch\_raw\_property; htrdr\_sky\_fetch\_svx\_property; htrdr\_sky\_trace\_ray;\\&&\\
                htrdr\_ground & Build scene and acceleration structure from input obj file describing the ground as a set of triangles, trace rays in the scene. & htrdr\_ground\_create; htrdr\_ground\_trace\_ray;\\&&\\
                htrdr\_sun & Implement a sun model, sample solar cone, access sun data. & htrdr\_sun\_create; htrdr\_sun\_sample\_direction; htrdr\_sun\_get\_radiance; \\&&\\
                htrdr\_camera & Implement a pinpoint camera model, trace a ray originating from the camera lens. & htrdr\_camera\_create; htrdr\_camera\_ray; \\&&\\

            \multicolumn{3}{l}{Most of the functions mentioned here can be found in the commented implementation of the renderer presented in \ref{part:images} \cite{starengine}.}\\
            \multicolumn{3}{l}{This list of functions is not comprehensive.}
	\end{tabularx}
        }
\end{sidewaystable}

\acknowledgments
Our many thanks go to F. Brient for providing us with the FIRE stratocumulus LES field, C. Strauss, D. Ricard and C. Lac for providing us with the 5m resolution congestus LES field, and C. Coustet for useful discussions. We acknowledge support from the Agence Nationale de la Recherche (ANR, grants HIGH-TUNE ANR-16-CE01-0010, http://www.umr-cnrm.fr/high-tune and MCG-RAD ANR-18-CE46-0012), from the french Programme National de Télédétection Spatiale (PNTS-2016-05), from Region Occitanie (Projet CLE-2016 EDStar) and from the French Minister of Higher Education, Research and Innovation for the PhD scholarship of the first author. The data and sources described in this paper are available at \url{https://www.meso-star.com/projects/high-tune/high-tune.html}. 
\newpage
\bibliography{all_ref}
\end{document}